\documentclass[12pt]{article}

\usepackage[utf8]{inputenc}      
\usepackage{amsmath, amssymb}    
\usepackage{graphicx}           
\usepackage{booktabs}             
\usepackage{hyperref}             
\usepackage{geometry}             
\usepackage{setspace}
\usepackage{apacite}
\usepackage{booktabs}
\usepackage[table]{xcolor}   
\definecolor{rowgray}{gray}{0.90} 
\usepackage{caption}
\usepackage{float}
\usepackage{enumitem}

\title{Identifying Unmeasured Confounders in Panel Causal Models:
A Two-Stage LM-Wald Approach
}
\author{Bang Quan Zheng}

\date{} 
\begin{document}

\maketitle

\begin{singlespace}
\begin{abstract}
\noindent Panel data are a cornerstone of causal inference in political science, psychology, and the social sciences more broadly. Yet these models often rest on the strong and rarely tested assumption of sequential ignorability—that no unmeasured variables jointly influence predictors and outcomes over time. This paper introduces the Two-Stage LM–Wald (2SLW) diagnostic, a general framework that extends Lagrange Multiplier and Wald tests to detect violations of this assumption in longitudinal causal models. The 2SLW procedure is demonstrated within a widely used class of panel models that separate stable differences between units from within-unit dynamics, allowing researchers to examine feedback processes over time. Through Monte Carlo simulations, we show that the 2SLW approach reliably identifies unmeasured confounding under three conditions: biased associations between variables, distortions of direct effects, and altered mediation pathways. An empirical illustration using real-world panel data demonstrates its practical application. Beyond offering a diagnostic tool, the 2SLW framework strengthens the theoretical and empirical foundations of causal inference in longitudinal research, enhancing the credibility of findings across studies of political behavior, attitude change, and policy feedback. The method is easily implemented in the \texttt{R} package \texttt{lavaan}.

\end{abstract}
\end{singlespace}

\vspace{0.25in}

\vspace{0.25in}

\begin{center}
\end{center}

\vspace{4in}

\doublespacing

\section{Introduction}

In recent decades, social scientists have increasingly turned to longitudinal panel data to uncover causal mechanisms and dynamic feedback processes. Approaches such as cross-lagged panel models (CLPMs), latent growth curve models, and random intercept cross-lagged panel models (RI-CLPMs) have become central tools for tracing temporal relationships among attitudes, preferences, and behaviors. These models inform theories of personality development, belief updating, and social learning—areas where causal processes unfold over time and experimental manipulation is often infeasible. Structural equation modeling (SEM) with latent variables has thus emerged as a key strategy for representing unobserved constructs and testing complex causal pathways. For instance, scholars have used longitudinal SEMs to examine links between partisanship and core values \cite{RN186}, authoritarianism and Republican support \cite{RN1115}, personality and political preferences \cite{bakker2021}, the ideological roots of moral intuition \cite{RN1106}, and the interplay between political and social trust \cite{RN1108}. Beyond political behavior, panel models are widely applied in comparative politics and international security to study dynamic systems in which outcomes and predictors evolve jointly—for example, conflict escalation and public support, or policy reform and regime legitimacy \cite{pluemper2005}. These applications illustrate a broader theoretical effort to model how individual-level dynamics shape aggregate political and social change.

Despite their promise, panel models rest on a critical but often implicit assumption: sequential ignorability, meaning that no unmeasured factors influence both the predictors and outcomes across time \cite{bellemare2017}. Violations of this assumption threaten not only statistical validity but also the theoretical interpretation of causal mechanisms, raising questions about whether observed dynamics reflect true feedback processes or hidden common causes. Prior methodological research shows that lagged explanatory variables does not resolve endogeneity because it requires the unrealistic assumption of no dynamics among unobserved confounders \cite{bellemare2017}. Similarly, although the recent development of the RI-CLPMs separates stable between-person differences from within-person dynamics, they do not eliminate bias from time-varying or dynamically persistent unobserved confounders. Thus, temporal ordering or cross-lagged paths alone do not guarantee exogeneity or valid causal interpretation. 

Even carefully specified panel models may conceal misspecification due to unmeasured stable factors, omitted cross-lagged paths, or residual dependencies among latent variables. Such hidden structure can bias estimates of autoregressive and cross-lagged effects, conflating true causal dynamics with spurious correlations. This problem is especially salient in studies of political behavior, where researchers examine how attitudes, beliefs, or behaviors evolve over time—for instance, whether political interest shapes later participation, media exposure influences attitudes across waves, or early experiences have lasting effects on engagement. By systematically examining locally misspecified parameters suggested by Lagrange Multiplier (LM) tests and validating them through Wald tests, researchers can detect traces of hidden confounding and distinguish genuine causal effects from distortions arising from omitted or misspecified model structure.

Sensitivity analysis provides an important framework for assessing potential violations of sequential ignorability \cite{RN1041, Blackwell_2014, RN818, RN827, Imai_Yamamoto_2013, LiuYamamoto2025}. However, most existing techniques are limited for multi-wave panel data with complex dependencies. They typically focus on one predictor at a time, ignore latent residual correlations, or cannot detect unmeasured confounders that influence multiple paths simultaneously \cite{harring2017}. Moreover, manual variable selection in sensitivity analysis introduces subjectivity, reducing reliability and limiting theoretical confidence in causal claims.  

To address these challenges, this study introduces the Two-Stage LM–Wald (2SLW) approach, a principled diagnostic designed to detect unmeasured confounding in panel causal models. Crucially, 2SLW does not recover the latent confounders themselves. Instead, it evaluates whether the restrictions implied by sequential ignorability—that is, the zero constraints on unmodeled paths and residual covariances—are statistically compatible with the observed data. LM tests assess local deviations from these restrictions, while Wald tests confirm whether detected deviations are substantively meaningful. Together, the two stages provide a rigorous diagnostic of violations of core identifying assumptions, enabling researchers to distinguish genuine causal dynamics from distortions caused by hidden confounding or model misspecification. 

Using Monte Carlo simulations within the RI-CLPM framework, we demonstrate that 2SLW reliably detects unmeasured confounders under biased correlations, distorted mediation pathways, and contamination of direct effects. An empirical application with real-world panel data further illustrates its practical value. Although demonstrated with RI-CLPMs, the 2SLW procedure generalizes naturally to a broad class of longitudinal designs, including standard CLPMs and higher-order RI-CLPMs. Together, these contributions position 2SLW as a practical and theoretically informed diagnostic that enhances the robustness of causal inference across disciplines by integrating psychometric theory, panel causal modeling, and sensitivity diagnostics.

\section{Panel Causal Inference \& Unmeasured Confounding}

Panel data are widely used in the social sciences to study dynamic relationships among attitudes, behaviors, and institutional outcomes. In political science, the most common approach is the CLPM. Despite differences in specification, CLPM-based designs rely on a common identifying assumption: conditional on observed variables and the modeled structure, no unmeasured factors jointly influence predictors and outcomes over time. Violations of this assumption are widely recognized as a central threat to causal inference in panel research \cite{bellemare2017} and may arise from omitted variables, model selection decisions, data limitations, or incorrect assumptions about error processes.

A central limitation of the standard CLPM is its treatment of unobserved heterogeneity. The model implicitly assumes that all relevant stable traits are either measured or substantively irrelevant. When this assumption fails, unobserved stable factors can confound cross-lagged estimates. In addition, unmodeled measurement error can bias autoregressive (AR) and cross-lagged (CL) parameters, potentially leading to misleading conclusions about within-person stability and change \cite{RN1101, RN692, RN1105, RN1109}.

\begin{figure}[H]
    \centering
    \caption{Diagrams of a 3-Wave RI-CLPM}
    \includegraphics[width=0.5\textwidth]{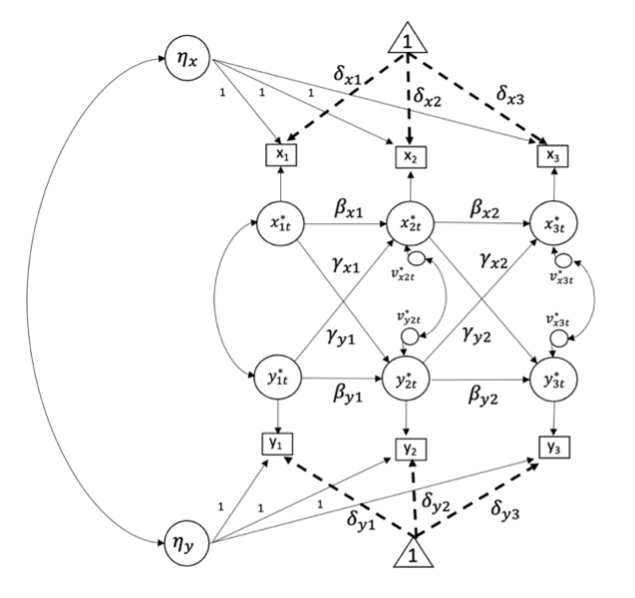}
    \label{fig:Figure_1}
\end{figure}

As Figure~\ref{fig:Figure_1} illustrates, the RI-CLPM extends the CLPM by separating stable between-person differences from within-person dynamics. By accounting for time-invariant unobserved heterogeneity through random intercepts, the RI-CLPM mitigates bias from stable confounders—one of the principal weaknesses of standard CLPMs \cite{RN692, RN689, RN1105}. However, this improvement does not address bias arising from time-varying unmeasured confounders that jointly affect predictors and outcomes across waves.

We use the RI-CLPM as an illustrative framework because it represents a demanding setting for longitudinal causal inference. Its ability to accommodate reciprocal and mediated relationships while disentangling stable heterogeneity from within-unit dynamics makes it attractive for studying feedback processes. At the same time, these features create multiple pathways through which unmeasured confounding can distort estimated effects. In this sense, the RI-CLPM serves as a stress test: a diagnostic that performs well in this setting is likely to generalize to simpler panel designs.

More broadly, approaches that model stable between-unit heterogeneity—such as fixed-effects models, dual-change models, or dynamic panel models—do not eliminate all sources of bias. Even when time-invariant confounders are absorbed, panel models remain vulnerable to time-varying unmeasured factors, including contextual shocks and evolving information environments. Moreover, lag structures cannot fully capture certain forms of serial dependence in disturbances, such as moving-average components, which induce structured residual correlations and bias parameter estimates when left unmodeled. From an identification perspective, such residual dependence is observationally equivalent to latent time-varying confounding, as both generate cross-equation correlations that violate sequential ignorability.

Together, these considerations highlight the limits of conventional panel specifications for diagnosing violations of identifying assumptions. Even sophisticated models rely on assumptions about unmeasured factors that are rarely testable using standard fit statistics or specification checks. This motivates the need for formal, model-based diagnostics capable of detecting unmeasured confounding in longitudinal causal models.

\subsection{From Static Measurement Models to Panel Causal Models}

This study extends the joint LM and Wald testing framework originally developed for confirmatory factor analysis (CFA) to longitudinal panel causal models. Although the statistical foundations of LM and Wald tests remain the same, the inferential environment in panel models differs fundamentally from that of static measurement models. These differences affect both the interpretation and behavior of modification diagnostics and motivate the development of the 2SLW procedure.

First, CFA evaluates a largely static covariance structure in which parameters such as loadings and residual correlations do not encode temporal ordering or causal direction. In contrast, panel causal models impose dynamic constraints: lagged paths represent directional effects over time, and random intercepts or related components separate stable heterogeneity from within-unit change. As a result, misspecification in panel models does not merely reflect measurement misfit but can signal violations of identifying assumptions about temporal processes and unmeasured confounding. Standard LM tests, when applied in this context, may therefore flag parameters whose misfit arises from dynamic dependence or omitted confounders rather than simple residual correlations.

Second, CFA applications typically use LM tests to detect localized departures from assumptions such as indicator independence. Panel models, however, feature feedback loops, lag structures, and serially dependent disturbances. Under these conditions, modification indices may reflect broader dynamic interactions rather than isolated parameter omissions. The 2SLW procedure addresses this issue by pairing the LM test with a Wald step that evaluates theoretically meaningful structural paths, helping distinguish superficial improvements in fit from evidence of deeper violations of the model’s causal structure.

To clarify the stochastic setting, consider a simple bivariate RI-CLPM. Let $x_{it}$ and $y_{it}$ denote measurements for individual $i$ at time $t$. Observed scores can be decomposed into between-person and within-person components,

\begin{align}
x_{it} &= \delta_{xt} + \eta_{xi} + x_{it}^*, \\
y_{it} &= \delta_{yt} + \eta_{yi} + y_{it}^*,
\end{align}

\noindent where $\eta_{xi}$ and $\eta_{yi}$ capture stable between-person differences and $x_{it}^*$ and $y_{it}^*$ represent within-person deviations. The within-person dynamics are modeled as

\begin{align}
x_{it}^* &= \beta_x x_{i(t-1)}^* + \gamma_x y_{i(t-1)}^* + v_{xit}, \\
y_{it}^* &= \beta_y y_{i(t-1)}^* + \gamma_y x_{i(t-1)}^* + v_{yit}.
\end{align}

\noindent Here, $\beta$ parameters represent autoregressive carry-over effects and $\gamma$ parameters represent cross-lagged influences. The disturbances $v_{xit}$ and $v_{yit}$ capture time-specific shocks not explained by the lag structure. A key identifying assumption is that these residuals are independent across equations,

\[
\rho = \mathrm{Cor}(v_{xit}, v_{yit}) = 0,
\]

\noindent which corresponds to sequential ignorability at the within-person level. When $\rho \neq 0$, an unmeasured time-varying factor jointly influences both variables, violating the causal interpretation of cross-lagged effects.

This feature is central to understanding the diagnostic role of the 2SLW procedure. In static CFA models, LM tests are typically interpreted as indicators of omitted covariance parameters and are often used to improve model fit. In panel causal models, however, significant LM suggestions may instead reflect violations of structural assumptions—such as the restriction \(\rho = 0\)—meaning that diagnostics can signal threats to identification rather than mere residual misfit. Taken together, 2SLW differs fundamentally from conventional modification-index--driven searches: it systematically screens for residual dependence, cross-validates findings via LM and Wald tests, and evaluates their substantive impact on focal parameters. In panel causal settings, this recasting of LM--Wald testing functions as a diagnostic for violations of identifying assumptions rather than a tool for fit improvement.

\section{Challenges in Sensitivity Analysis}

Sensitivity analysis is a central strategy for evaluating the robustness of causal conclusions to unmeasured confounding. Political methodologists have developed formal tools to assess how inferences might change under hypothetical violations of identifying assumptions \cite{RN1041, Blackwell_2014, RN818, RN827, LiuYamamoto2025}. However, most existing approaches are difficult to extend to complex longitudinal settings such as panel models with reciprocal relationships and repeated measurements. In these contexts, confounding may arise from both time-invariant and time-varying unobserved factors that influence multiple variables across waves, creating bias structures that are more intricate than those typically considered in static treatment–outcome frameworks.

Recent advances in panel causal analysis address time-varying confounding using Bayesian sensitivity analysis and latent factor approaches \cite{LiuYamamoto2025}. These methods provide probabilistic assessments of how strongly unmeasured confounders would need to operate to alter conclusions. Yet they generally focus on quantifying the magnitude of potential bias rather than identifying where in a structural model confounding is most likely to occur. Researchers must specify hypothetical confounder–outcome relationships and explore high-dimensional sensitivity-parameter spaces, often at considerable computational and interpretive cost.

The 2SLW approach serves a distinct purpose. Rather than estimating how large unobserved confounding would have to be, it functions as a diagnostic tool that evaluates whether the observed data exhibit patterns inconsistent with the maintained structural assumptions. Specifically, 2SLW identifies particular paths or constraints in a longitudinal model that generate empirical misfit, thereby localizing potential sources of bias. In this sense, 2SLW complements rather than replaces sensitivity analysis: it helps determine where identifying assumptions may fail before researchers assess how much such violations would matter.

This diagnostic role becomes especially important in panel models with lagged variables and feedback processes. Even when baseline treatments are randomized, subsequent lagged predictors may be endogenous because they are shaped by prior outcomes, stable traits, or evolving contextual forces that are rarely observed \cite{bellemare2017}. In such settings, confounding does not enter through a single omitted variable but propagates dynamically across multiple time points and pathways. Conventional sensitivity analyses—typically developed for static or single-equation models—are ill-equipped to detect these distributed forms of misspecification. A localized diagnostic framework is therefore necessary to identify where longitudinal identifying restrictions break down.

\subsection{Research Strategy}

The illustration of the 2SLW method proceeds in three steps. First, drawing on a covariance structure-based data-generating process, we simulate data from a population RI-CLPM that includes omitted, unmeasured confounders. We then fit an analysis model that intentionally excludes these confounders, mimicking the types of misspecification researchers may encounter in applied work. This deliberate violation of the sequential ignorability assumption induces model misfit, thereby creating a setting in which the ability of the 2SLW method to detect misspecification can be evaluated. Second, we apply a univariate LM test to identify the top 25 candidate parameters ranked by LM $\chi^2$ statistics and Expected Parameter Change (EPC) values, excluding those with small EPCs, violations of temporal ordering, or potential non-identification problems in the RI-CLPM. Third, we refine the model using a forward stepwise Wald test, retaining only parameters with $p<0.05$ while removing any that cause convergence issues, non-positive definite covariance matrices, or negative variances. Identified parameters suggest possible unmeasured confounders, which we incorporate into an extended RI-CLPM to assess their effect on autoregressive and cross-lagged coefficients. Substantial changes imply bias from unmeasured confounding, whereas stability increases confidence in the model’s robustness. This procedure, combining LM-driven detection with Wald-based parsimony, offers a structured, scalable approach to sensitivity analysis for SEM-based panel models and can be implemented in \texttt{R} using \texttt{lavaan}.

\section{LM and Wald Tests as the Statistical Basis of 2SLW}

This section formalizes the statistical foundation of the 2SLW approach. The method relies on local tests of model misspecification that evaluate whether constrained parameters—interpreted as absent confounding or omitted paths—are statistically distinguishable from zero. These diagnostics are operationalized using the LM test and the Wald test, which provide complementary perspectives on whether relaxing structural constraints would meaningfully alter model fit and substantive parameter estimates.

To formalize the estimation problem, consider the maximum likelihood discrepancy function used in latent variable modeling:

\begin{equation}
F(\boldsymbol{\theta}) = \log \left| \boldsymbol{\Sigma}(\boldsymbol{\theta}) \right| - \log \left| \mathbf{S} \right|
+ \operatorname{tr}\!\left( \mathbf{S} \, \boldsymbol{\Sigma}(\boldsymbol{\theta})^{-1} \right) - p,
\label{eq:7}
\end{equation}

\noindent where $\mathbf{S}$ denotes the sample covariance matrix and $\boldsymbol{\Sigma}(\boldsymbol{\theta})$ is the model-implied covariance matrix as a function of the parameter vector $\boldsymbol{\theta}$. Let $\hat{\boldsymbol{\theta}}$ represent the vector of constrained estimators of $\boldsymbol{\theta}$ under the constraints $\mathbf{h}(\boldsymbol{\theta}) = 0$, which encode restrictions such as fixed factor loadings or covariate effects. This formulation provides a foundation for deriving bias corrections in regression models with latent variables, particularly when covariates induce distortions in parameter estimates.
 With $r$ constraints, the constraint vector is $\mathbf{h}(\boldsymbol{\theta})' = (h_1, \dots, h_r)$. Minimizing $F(\boldsymbol{\theta})$ under these constraints yields derivative matrices
$\mathbf{g} = \partial F / \partial \boldsymbol{\theta}$ and $\mathbf{L}' = \partial \mathbf{h} / \partial \boldsymbol{\theta}$, along with a vector of Lagrange multipliers $\hat{\boldsymbol{\lambda}}$ satisfying

\begin{equation}
\hat{\mathbf{g}} + \hat{\mathbf{L}}' \hat{\boldsymbol{\lambda}} = 0, \quad \mathbf{h}(\hat{\boldsymbol{\theta}}) = 0.
\label{eq:8}
\end{equation}

\noindent The asymptotic covariance matrix of the estimated parameters can be expressed in block form as

\begin{equation}
\begin{bmatrix} 
\mathbf{H} & \mathbf{L}' \\ 
\mathbf{L} & \mathbf{O} 
\end{bmatrix}^{-1} =
\begin{bmatrix} 
\mathbf{H}^{-1} - \mathbf{H}^{-1}\mathbf{L}'(\mathbf{L}\mathbf{H}^{-1}\mathbf{L}')^{-1} & 
\mathbf{H}^{-1} \mathbf{L}'(\mathbf{L}\mathbf{H}^{-1}\mathbf{L}')^{-1} \\ 
(\mathbf{L}\mathbf{H}^{-1}\mathbf{L}')^{-1} \mathbf{L}\mathbf{H}^{-1} & 
-(\mathbf{L}\mathbf{H}^{-1}\mathbf{L}')^{-1} 
\end{bmatrix} =
\begin{bmatrix} 
\mathbf{M} & \mathbf{T}' \\ 
\mathbf{T} & -\mathbf{R} 
\end{bmatrix}.
\label{eq:9}
\end{equation}

\subsection{The LM Test and Expected Parameter Change (EPC)}

The LM test evaluates whether the score vector associated with constrained parameters is nonzero at the boundary of the parameter space, providing a local assessment of model misspecification. For a single constraint, the univariate LM statistic is

\begin{equation}
T_{\text{LM},i} = n \, \hat{\lambda}_i^2 \, \hat{\mathbf{R}}_{ii}^{-1} \sim \chi^2_1.
\label{eq:10}
\end{equation}
The LM test is based on the score vector evaluated at the constrained estimates. 
A first-order Taylor expansion of the likelihood around the constrained solution 
implies that the change in a parameter after relaxing a constraint can be approximated by

\begin{equation}
\Delta \boldsymbol{\theta} \approx \mathbf{I}^{-1}\mathbf{s},
\label{eq:epc}
\end{equation}

\noindent where $\mathbf{s}$ is the score vector and $\mathbf{I}$ is the information matrix. 
For a single constrained parameter $\theta_i$, this yields the Expected Parameter Change (EPC):

\begin{equation}
\mathrm{EPC}_i = \frac{\hat{\lambda}_i}{\hat{\mathbf{R}}_{ii}},
\label{eq:11}
\end{equation}

\noindent where $\hat{\lambda}_i$ is the LM component associated with parameter 
$\theta_i$, and $\hat{\mathbf{R}}_{ii}$ is the corresponding diagonal element of the 
information matrix. The EPC therefore provides a first-order approximation of how 
much the parameter would change if the constraint were released. Large EPC values 
indicate potential overconstraints and help identify locally omitted paths or 
unmeasured confounders.

Under local alternatives, the unrestricted estimator of the constrained parameter $\boldsymbol{\theta}_r$ can be approximated using the Lagrange multiplier:

\begin{equation}
\hat{\boldsymbol{\theta}}_r \;\approx\;
\hat{\mathbf{H}}_{rr}^{-1}\hat{\mathbf{L}}_r^{\prime}\hat{\boldsymbol{\lambda}},
\end{equation}

\noindent where $\hat{\mathbf{H}}_{rr}$ is the relevant block of the Hessian and $\hat{\mathbf{L}}_r$ is the constraint Jacobian. This shows that parameters flagged by the LM test (nonzero $\hat{\boldsymbol{\lambda}}$) are precisely those expected to attain nonzero values once freed.

\subsection{The Wald Test}

The Wald test complements the LM test by evaluating the magnitude and statistical significance of parameters once they are unconstrained. For a vector of candidate parameters $\hat{\boldsymbol{\theta}}_r$, the Wald statistic is
\begin{equation}
W = n\, \hat{\boldsymbol{\theta}}_r' 
\big(\hat{\mathbf{L}}_r \hat{\mathbf{H}}_{rr}^{-1} \hat{\mathbf{L}}_r'\big)^{-1} 
\hat{\boldsymbol{\theta}}_r \sim \chi^2_r,
\label{eq:12}
\end{equation}
where $\hat{\mathbf{H}}_{rr}$ is the estimated information matrix and $\hat{\mathbf{L}}_r$ encodes any linear constraints.

For a single parameter, this reduces to
\begin{equation}
W_i = \frac{\hat{\theta}_i^2}{\mathrm{Var}(\hat{\theta}_i)} \sim \chi^2_1,
\end{equation}
where $\mathrm{Var}(\hat{\theta}_i)$ is typically the $i$th diagonal element of $\hat{\mathbf{H}}_{rr}^{-1}$. In the 2SLW framework, the Wald test is applied sequentially to parameters flagged by the LM test, confirming which constraints correspond to substantively meaningful omissions.

The LM and Wald tests are theoretically equivalent in that they tend to identify the same parameters as potentially misspecified or omitted. For the same candidate parameter $\hat{\theta}_i$, the LM statistic is $T_{{\rm LM},i}$. Although $W_i$ and $T_{{\rm LM},i}$ are generally not numerically equal, both highlight the same constraints that may be overly restrictive, signaling potentially omitted paths or unmeasured confounders. In this sense, the two tests provide complementary but consistent information about model misspecification.

\subsection{Complementarity of LM and Wald Tests}

Under standard regularity conditions, the LM test and Wald test evaluate distinct but complementary aspects of constrained parameters: LM assesses local necessity at the boundary, and Wald test assesses global estimability. Neither test alone is sufficient for diagnosing structurally omitted paths.

The 2SLW procedure operationalizes this complementarity by first using LM to screen for candidate misfit parameters, then applying the Wald test to confirm their substantive impact once unconstrained. A parameter is considered a meaningful structural omission if and only if it satisfies both: 
(i) local necessity (LM significant), and 
(ii) global estimability (Wald significant). This sequential application provides a rigorous and mathematically justified method for identifying omitted confounders in RI-CLPMs or other structural models.

\section{Monte Carlo Simulation}
In this section, we evaluate the 2SLW approach using Monte Carlo simulations. The simulation study consists of three main steps. In the first step, we generate simulated data from three distinct RI-CLPMs, which serve as the population models. The data generation follows the RI-CLPM covariance matrix structure (see Appendix). Specifically, a $p$-dimensional variable, $\mathbf{z}_i$, is generated as

\begin{equation}
\mathbf{z}_i = \boldsymbol{\Sigma(\theta)}^{1/2} \boldsymbol{\varepsilon}_i,
\tag{14}
\end{equation}

\noindent where $\boldsymbol{\Sigma(\theta)}$ is the covariance matrix. The vector $\boldsymbol{\varepsilon}_i \sim N(\mathbf{0}, \mathbf{I})$ represents random errors drawn from a standard multivariate normal distribution with zero mean and identity variance. For the parameter values, we set AR parameters $\beta_{xi}$ and CL parameters $\gamma_{yi}$ at 0.25 and 0.15, respectively. We use the same parameter values and specifications for the time-varying component of the RI-CLPM across all models to ensure comparability and validate the simulation results. Although parameter choices in simulations can be arbitrary, they do not affect test statistics or fit indices when models are correctly specified. 

In the second step, we specify three analysis models that intentionally omit confounders affecting both the independent and outcome variables. These misspecified models are expected to exhibit poor fit and biased parameter estimates. In the third step, we apply both the LM and Wald tests to identify the omitted confounders and evaluate their relative effectiveness. All simulation analyses are conducted in \texttt{R} using the \texttt{lavaan} package \cite{RN547}.

\subsection{Population Models}

Each population model is a four-wave RI-CLPM with two latent variables, $X$ and $Y$, measured by single indicators at each time point. 
The RI-CLPM structure separates within-person fluctuations from stable between-person differences, such as deviations from one’s mean trajectory. To increase complexity while avoiding overfitting, we use four time points rather than adding more indicators per construct.

\section{Population Models \& Specification}

\begin{figure}[H]
    \centering
    \caption{Diagrams of Population Models}

    \begin{minipage}{0.5\textwidth}
        \centering
        \textbf{Model 1}\\[0.5ex]
        \includegraphics[width=\linewidth]{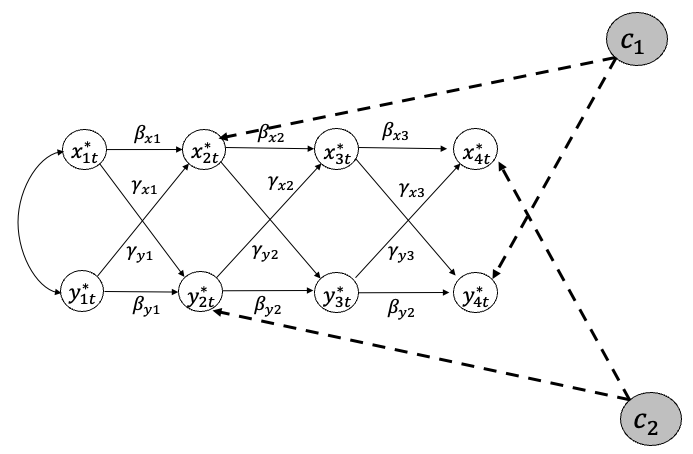}
    \end{minipage}
    \hfill
    \begin{minipage}{0.45\textwidth}
        \centering
        \textbf{Model 2}\\[0.5ex]
        \includegraphics[width=\linewidth]{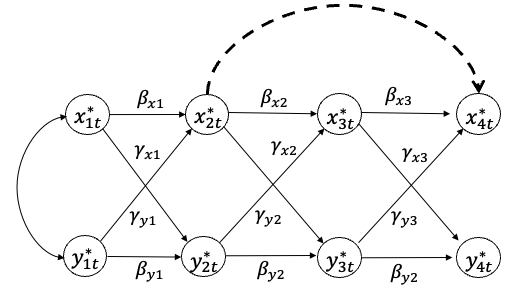}
    \end{minipage}

    \vspace{1em}

    \begin{minipage}{0.45\textwidth}
        \centering
        \textbf{Model 3}\\[0.5ex]
        \includegraphics[width=\linewidth]{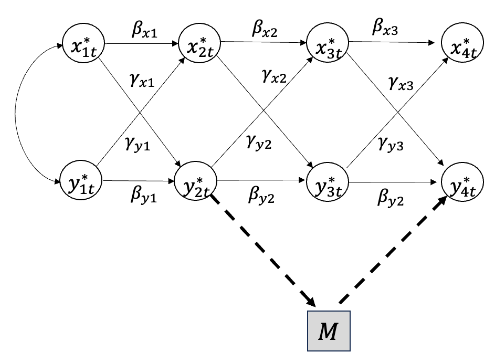}
    \end{minipage}

    \caption*{\footnotesize Note: To simplify the notations, residuals are not shown in the diagrams.}
    \label{fig:Figure2}
\end{figure}

Figure~\ref{fig:Figure2} presents diagrams of three RI-CLPMs that share a common baseline structure but differ in their unmeasured confounding designs. To simplify notation, the remainder of the paper uses $WFX$ to denote the autoregressive path and $WFY$ the cross-lagged path in the within-factor component, while $BX$ and $BY$ represent the between-factor component. This convention makes the statistical output easier to present. As illustrated in Figure~\ref{fig:Figure2}, each model is based on a four-wave RI-CLPM framework that includes both within-person and between-factor components. For the within-factor component, two sets of latent variables, $WFX_i$ and $WFY_i$, for $i=1,2,\dots,4$, are specified. Each latent variable is measured by a single observed indicator, $x_i$ and $y_i$, respectively. For the between-person component, two latent factors, $BX$ and $BY$, are introduced to capture stable individual traits. Additionally, $BX$ and $BY$ are allowed to correlate, reflecting the association between stable traits across constructs. 

\subsection{Model Variants}

As shown in Figure~\ref{fig:Figure2} , Model 1 introduces two higher-order latent confounders, $c_1$ and $c_2$, representing dynamic confounding between the residuals of $WFX_2$ and $WFY_4$, and between the residuals of $WFY_2$ and $WFX_4$, respectively. For example, in political communication, $c_1$ could capture unobserved partisan media exposure that simultaneously influences political efficacy and later policy support, while $c_2$ might reflect unmeasured interpersonal political discussion affecting earlier policy support and later efficacy. By inducing correlated residual variation across time and constructs, these confounders can bias cross-lagged path estimates.

Model 2 imposes a unidirectional direct path from $WFX_2$ to $WFX_4$, representing a case where an earlier latent variable exerts a causal influence on a later one while bypassing intermediate processes. In political behavior terms, this could capture citizens’ trust in government at time $t$ directly shaping trust at $t+2$, without accounting for mediating influences such as media exposure or political events. This specification violates temporal stationarity by assuming the effect remains constant across time, thereby conflating transient causal effects with stable dispositions or prior attitudes.

Model 3 incorporates unmeasured confounding within a mediation framework, where the residual of $WFY_2$ on the residual of $WFY_4$ is mediated indirectly through the residual of an unobserved mediator $M$, which itself is regressed on $x_1$. For example, political engagement could mediate the relationship between political efficacy at two time points. Omitting $M$ risks misattributing mediated effects as direct effects, thereby biasing causal interpretations.  

In sum, all three models intentionally violate the sequential ignorability assumption. Simulated data from these population models are analyzed using deliberately misspecified models that omit latent confounders or structural paths, generating the expected model misfit. The simulation examines correlations and regressions among latent variables while fixing factor loadings at 1, ensuring that observed changes reflect only time-varying effects. Because the factor loadings are fixed, between-person correlations and residual variances are excluded. To evaluate generalizability, additional analyses of a five-wave RI-CLPM with two indicators, as well as a conventional CLPM, are presented in the Appendix.

\subsection{Baseline Model Specification and Validity}

Parameter estimation, model fit, and sensitivity analysis all depend on correct model specification. Before conducting simulations on the three variants, it is crucial to verify that both the baseline population model and the analysis model meet the necessary asymptotic assumptions—that is, how an estimator behaves as the sample size approaches infinity, typically converging in probability to the true parameter, thus providing a theoretical basis for inference when finite-sample behavior is uncertain. 

To achieve this, we evaluate the model fit across repeated samples with varying sample sizes. If a model is identified and adequately specified, we expect the model to follow a standard $\chi^2$ distribution, $T_{\mathrm{ML}} \sim \chi^2_{\mathrm{df}}$, as sample size grows larger, demonstrating asymptotic properties \cite{RN882, RN664}. Accordingly, we use Monte Carlo simulation to generate data from the predefined population models as illustrated in Figure~\ref{fig:Figure2} and estimate the corresponding analysis models using the simulated data.

\begin{table}[htbp]
\centering
\caption{Simulation results of model fit statistics across sample sizes.}
\begin{tabular}{rrrrrrrrr}
\toprule
N & $\chi^2$ & SD & $P$-value & Rej Rate & NFI & CFI & RMSEA \\
\midrule
100    & 9.214 & 4.402 & 0.490 & 0.068 & 0.937 & 0.959 & 0.119 \\
200    & 9.195 & 4.170 & 0.486 & 0.058 & 0.985 & 1.000 & 0.000 \\
300    & 9.178 & 4.381 & 0.491 & 0.050 & 0.991 & 1.000 & 0.000 \\
500    & 8.909 & 4.472 & 0.513 & 0.056 & 0.991 & 0.998 & 0.028 \\
800    & 9.034 & 4.380 & 0.501 & 0.060 & 0.997 & 1.000 & 0.000 \\
1000   & 9.018 & 4.140 & 0.495 & 0.042 & 0.999 & 1.000 & 0.000 \\
5000   & 9.138 & 4.116 & 0.484 & 0.040 & 1.000 & 1.000 & 0.000 \\
10000  & 8.921 & 4.546 & 0.513 & 0.050 & 1.000 & 1.000 & 0.006 \\
\bottomrule
\end{tabular}
\label{tab:table_1}
\end{table}

 We conducted 500 trials and calculated the average statistics, which are reported in Table~\ref{tab:table_1}. We examined the performance of the analysis model by varying the sample sizes from 100 to 10,000. Since the simulated data were normally distributed, the maximum likelihood estimator was sufficient to examine the basic statistical performance and asymptotic properties. 

Table~\ref{tab:table_1} presents the average $\chi^2$ test statistics, standard deviations, $P$-values, rejection rates, and fit indices (NFI, CFI, RMSEA) by sample size. As shown in Table~\ref{tab:table_1}, as the sample size increases, all mean test statistics approach the expected values: $\chi^2 = 9$, $\mathrm{SD} = 4.24$, $p$-value = 0.50, and empirical rejection rate is $0.05$. In addition, NFI and CFI approach $1$, and RMSEA approaches $0$ when the sample sizes are greater than 200. Naturally, with smaller sample sizes, test statistics and fit indices deviate from their expected values—a well-documented issue for the maximum likelihood estimator \cite{RN708, RN949, RN1095}. The collective statistical indicators demonstrate that the population and baseline analysis models are well-specified. Based on this baseline model, we will proceed with various analyses.

\section{The Two-Stage Process for Confounder Searches}

\subsection{Stage One: Selection of Testing Parameters}

We start with a univariate LM test to identify the top candidate parameters based on their LM $\chi^2$ statistics and EPC values. From this test, we retain the top 25 candidate parameters\footnote{The number of parameters retained is guided by model complexity. We adopt 25 as a conservative cutoff to ensure that substantively important confounding or omitted parameters are unlikely to be excluded.} and exclude any that violate the following criteria:

\begin{singlespace}
\begin{enumerate}[label=(\roman*)]
    \item Parameters that contradict temporal ordering (e.g., a variable at time $t$ predicting one at $t-1$) or conflict with the population model (e.g., LM-suggested correlations between autoregressively linked latent variables or fixed variances). In RI-CLPMs, suggestions linking latent variables to their indicators, which may cause non-identification or convergence problems, are also excluded.
    \item Parameters that introduce redundant or meaningless relations, such as correlations between autoregressive and cross-lagged paths, are removed.
    \item Parameters with extremely small EPC values are excluded.
    \item Finally, we add the remaining candidate parameters one at a time to the analysis model; any parameter that causes convergence or identification issues is removed.
\end{enumerate}
\end{singlespace}

\begin{doublespace}
\subsection{Stage Two: The Wald Test}

The Wald test follows, assessing candidate parameters identified by the LM test through a forward stepwise procedure. Each parameter is added individually, and the Wald statistic and $p$-value are computed at each step. This process continues until all candidates are evaluated, enhancing model parsimony and reducing false positives from overfitting. Ultimately, only parameters with $p$-values below 0.05 are retained for further analysis. 

\subsection{Simulation Result}

Tables 2--4 present the Monte Carlo simulation results, where the gray-highlighted entries indicate cases in which the confounding variables match those in the population models. In Model 1 (Table 2), the residual correlations $WFX_{4} \sim\!\!\sim WFY_{2}$ and $WFX_{2} \sim\!\!\sim WFY_{4}$ were successfully identified by the LM chi-square, EPC, and Wald tests, indicating that these variables share common variance driven by a latent confounder. Notably, while the path $WFX_{4} \sim WFY_{2}$ was flagged by the LM chi-square with a strong EPC, its Wald test statistic was small (0.537) and statistically insignificant. In Model 2 (Table 3), the LM test identified several direct effects, including $WFX_{4} \sim WFX_{2}$. However, only this path was validated by the Wald test ($p = 0.019$), consistent with the population model.  In Model 3 (Table 4), the LM test and EPC flagged multiple paths, including $WFY_{4} \sim M$ and $WFY_{4} \sim WFY_{2}$. Of these, only these two paths were further validated by the Wald test ($p < 0.05$), again aligning with the population model.  

Overall, the LM tests effectively detected confounders through large chi-square statistics and EPC values, while the Wald tests confirmed true effects and rejected false ones. Moreover, the estimated AR and CL paths closely matched the population parameters (see Tables A1--A3 in the Appendix). Additional simulations with more complex models, including higher-order latent variables, additional waves, and multiple indicators, are presented in the Appendix and show similarly robust performance.

\renewcommand{\arraystretch}{0.75}
\begin{table}[H]
\centering
\caption{Model 1 (Correlation) Simulation Result}
\label{tab:table_2}
\begin{tabular}{llcccc}
\toprule
 &  & \textbf{LM $\chi^2$} & \textbf{EPC} & \textbf{Wald Test} & $P$-value \\
\midrule
\rowcolor{rowgray} WFX$_4$ $\sim\sim$ & WFY$_2$ & 575.299 & 0.447 & 12.780 & 0.000 \\
WFX$_4$ $\sim$     & WFY$_2$ & 561.970 & 0.810 & 0.537 & 0.464 \\
\rowcolor{rowgray} WFX$_2$ $\sim\sim$ & WFY$_4$ & 540.303 & 0.419 & 7.874 & 0.005 \\
WFY$_4$ $\sim$     & WFX$_2$ & 515.640 & 0.764 & 0.372 & 0.796 \\
WFY$_4$ $\sim$     & X$_2$   & 499.933 & 0.642 & 0.481 & 0.488 \\
\bottomrule
\end{tabular}
    \caption*{\footnotesize Note: We follow the same notation conventions used in \texttt{lavaan}, where `$=\sim$' denotes a latent variable definition, `$\sim$' indicates a regression relationship, and `$\sim\sim$' represents a covariance or correlation.}
\end{table}

\renewcommand{\arraystretch}{0.75}
\begin{table}[H]
\centering
\caption{Model 2 (Direct Effect) Simulation Result}
\label{tab:table_3}
\begin{tabular}{llcccc}
\toprule
 &  & \textbf{LM $\chi^2$} & \textbf{EPC} & \textbf{Wald Test} & $P$-value \\
\midrule
\rowcolor{rowgray} WFX$_4$ $\sim$     & WFX$_2$ & 222.566 & 1.142  & 5.547 & 0.019 \\
WFX$_4$ $\sim\sim$ & WFY$_3$ & 196.282 & -0.110 & 0.528 & 0.467 \\
WFX$_4$ $\sim$     & Y$_3$   & 190.300 & -1.024 & 1.278 & 0.258 \\
WFX$_4$ $\sim$     & X$_2$   & 172.917 & 0.458  & 0.049 & 0.824 \\
\bottomrule
\end{tabular}
\end{table}

\renewcommand{\arraystretch}{0.75}
\begin{table}[H]
\centering
\caption{Model 3 (Mediation) Simulation Result}
\label{tab:table_4}
\begin{tabular}{lllll}
\toprule
 & LM & EPC & Wald Test & $P$-value \\ 
\midrule
\rowcolor{rowgray} WFY$_4$ $\sim$ M        & 285.407 & 0.742  & 20.932 & 0.000     \\
\rowcolor{rowgray} WFY$_4$ $\sim$ WFY$_2$  & 183.201 & 1.578  & 5.117  & 0.024 \\
WFY$_4$ $\sim$ Y$_3$                        & 148.494 & 1.367  & 0.546  & 0.460 \\
WFY$_1$ $\sim$ X$_1$                        & 143.828 & -2.882 & 0.115  & 0.735 \\
WFY$_2$ $\sim\sim$ WFY$_4$                 & 138.537 & 0.079  & 0.065  & 0.799 \\
WFY$_4$ $\sim$ Y$_2$                        & 132.760 & 0.447  & 3.844  & 0.050 \\
WFY$_4$ $\sim$ X$_1$                        & 116.008 & 0.416  & 0.387  & 0.534 \\ 
\bottomrule
\end{tabular}
\end{table}
\renewcommand{\arraystretch}{1.0} 

\section{Empirical Application}

This section provides an empirical illustration of the 2SLW approach within a RI-CLPM. The goal is twofold: to demonstrate how 2SLW can be implemented in a longitudinal SEM framework and to illustrate how RI-CLPM can be used to study political attitudes, where its adoption remains relatively limited.

Following the framework of \citeA{bakker2021}, who examine the relationship between Big Five personality traits and political preferences, we construct a three-wave panel using data from 2008, 2014, and 2020. While their study primarily investigates broad personality–politics associations, our illustration focuses specifically on agreeableness and immigration policy attitudes, providing a substantive extension. To better capture dynamic within-person processes, we apply a RI‑CLPM, which separates stable between-person differences from within-person fluctuations over time. This allows us to examine how changes in agreeableness relate to changes in immigration attitudes, while accounting for stable individual heterogeneity. In this way, our application both adapts the conceptual framework of prior work and demonstrates the practical implementation of RI‑CLPM in political science contexts.

We use data from the Longitudinal Internet Studies for the Social Sciences (LISS) panel, a long-running internet-based survey of individuals and households in the Netherlands administered by CentERdata. Agreeableness is measured using two items: (1) Interested in people'' and (2) I am not interested in other people’s problems'' (reverse coded). Immigration attitudes are assessed with three items: (1) There are too many people of foreign origin or descent in the Netherlands,'' (2) Legally residing foreigners should be entitled to the same social security as Dutch citizens,'' and (3) It should be made easier to obtain asylum in the Netherlands.'' Responses are recorded on a five-point Likert scale ranging from fully disagree'' to ``fully agree.''

Our analytic strategy proceeds in two steps. First, we estimate a baseline RI-CLPM to assess the dynamic associations between agreeableness and immigration attitudes. Second, we apply the 2SLW procedure as a diagnostic tool to identify latent residual structures that may indicate unmodeled dependencies or potential sources of bias. We then compare model fit, autoregressive (AR) coefficients, and cross-lagged (CL) estimates between the baseline and improved models.

\begin{figure}[H]
    \centering
    \caption{Diagram of the RI-CLPM}
    \includegraphics[width=0.5\textwidth]{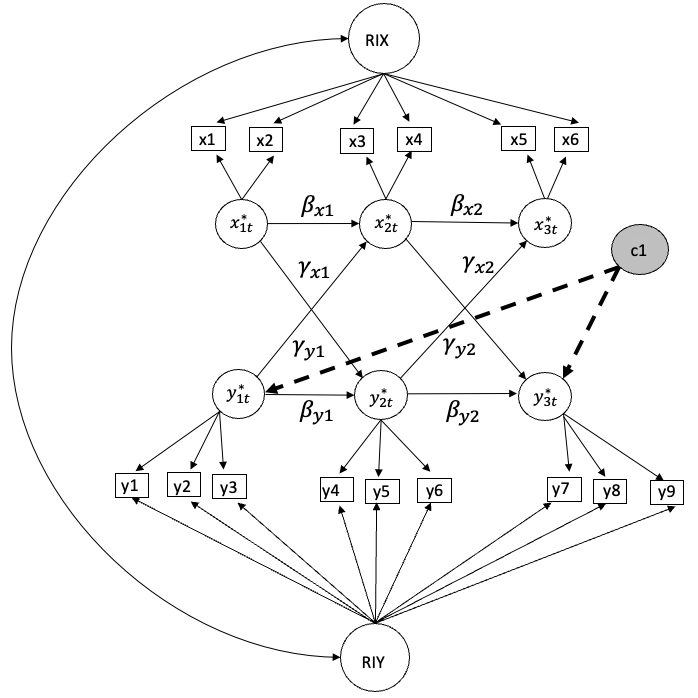} 
    \label{fig:Figure_4}
\end{figure}. 

The agreeableness and immigration-attitude items are denoted as $X_i$ and $Y_i$ in Figure~\ref{fig:Figure_4}. Latent within-person factors WFX$_i$ and WFY$_i$ are constructed for each wave. Undirected paths represent autoregressive effects, while directed paths represent cross-lagged effects. Dashed paths from latent factor $c_1$ to WFY$_1$ and WFY$_3$ indicate residual structure identified by the 2SLW procedure.

The latent factor \(c_1\) captures residual covariance between WFY\(_1\) and WFY\(_3\). Importantly, we do not interpret \(c_1\) as a confirmed causal confounder. Rather, it serves as a diagnostic representation of unmeasured heterogeneity, model misspecification, or other latent structure. Incorporating \(c_1\) allows us to evaluate the robustness of cross-lagged estimates and assess whether the baseline specification omitted substantively relevant structure. The 2SLW procedure identifies statistically detectable residual dependence consistent with omitted structure; it does not distinguish between omitted causes, measurement artifacts, or dynamic misspecification.

Table~\ref{tab:lm_wald_results} reports LM and Wald tests. Four parameters were flagged by the LM test, of which two are supported by the Wald test. The residual association between WFY$_1$ and WFY$_3$ shows the strongest statistical evidence and is incorporated into the improved model.

\renewcommand{\arraystretch}{0.75}
\begin{table}[H]
\centering
\caption{The 2SLW Test Result}
\begin{tabular}{lllcccc}
\hline
       &        &                       & \textbf{LM $\chi^2$} & \textbf{EPC} & \textbf{Wald Test} & \textit{P}-value \\
\hline
\rowcolor[gray]{0.9}
WFY$_1$ & $\sim\sim$ & WFY$_3$                  & 53.615  & 0.143  & 15.464 & 0.000 \\
WFY$_3$ & $=\sim$    & Immigrant Benefits (08) & 43.627  & 0.679  &  1.596 & 0.207 \\
WFX$_1$ & $\sim\sim$ & WFY$_3$                  & 40.390  & -0.244 &  0.593 & 0.441 \\
\rowcolor[gray]{0.9}
WFY$_2$ & $=\sim$    & Immigrant Benefits (08) & 38.387  & 0.969  &  4.862 & 0.027 \\
\hline
\end{tabular}
\label{tab:lm_wald_results}
\end{table}

Table~\ref{tab:lm_wald_results} presents the results of the LM and Wald tests. The LM test identified four parameters, of which only two were statistically significant according to the Wald test, as highlighted. Notably, the correlation between WFY$_1$ and WFY$_3$ yielded the largest $\chi^2$ value and the highest Wald test statistic, indicating strong statistical significance. These two parameters will be incorporated into the existing model.

\renewcommand{\arraystretch}{0.75}
\begin{table}[ht]
\centering
\caption{AR \& CL Coefficients of Original and Improved Models}
\begin{tabular}{l c c c c c c c}
\toprule
 & \multicolumn{3}{c}{Original Model} & \multicolumn{3}{c}{Improved Model} & {Difference} \\
\cmidrule(r){2-4} \cmidrule(r){5-7}
Variable & {Estimate} & {Std. Err} & $P$-value & {Estimate} & {Std. Err} & $P$-value & {Diff} \\
\midrule
\multicolumn{8}{l}{WFX$_2$ $\sim$} \\
WFX$_1$ & 0.824 & 0.031 & 0.000 & 0.825 & 0.027 & 0.000 & 0.001 \\
WFY$_1$ & 0.121 & 0.160 & 0.450 & -0.096 & 0.016 & 0.010 & -0.217 \\
\midrule
\multicolumn{8}{l}{WFY$_2$ $\sim$} \\
WFY$_1$ & 1.585 & 0.185 & 0.000 & 0.977 & 0.023 & 0.000 & -0.608 \\
WFX$_1$ & 0.047 & 0.024 & 0.049 & -0.025 & 0.025 & 0.026 & -0.072 \\
\midrule
\multicolumn{8}{l}{WFX$_3$ $\sim$} \\
WFX$_2$ & 0.961 & 0.030 & 0.000 & 0.906 & 0.029 & 0.000 & -0.055 \\
WFY$_2$ & 0.089 & 0.081 & 0.268 & 0.006 & 0.015 & 0.854 & -0.083 \\
\midrule
\multicolumn{8}{l}{WFY$_3$ $\sim$} \\
WFY$_2$ & 1.015 & 0.076 & 0.000 & 0.861 & 0.071 & 0.000 & -0.154 \\
WFX$_2$ & -0.071 & 0.021 & 0.001 & -0.051 & 0.081 & 0.136 & 0.020 \\

\midrule
\multicolumn{8}{l}{\textit{$C_1$} $\sim$} \\
WFY$_1$ & & & & 0.584 & 0.112 & 0.000 &  \\
WFY$_3$ & & & & 0.180 & 0.143 & 0.028 & \\
\midrule

\multicolumn{8}{l}{Model Fit} \\
$\chi^2$ & 679.509 & & & 130.809 & & & -548.700 \\
$\mathit{df}$  & 74 & & & 65 & & & -9.000 \\
CFI & 0.991 & & & 1.000 & & & 0.009 \\
NFI & 0.990 & & & 0.998 & & & 0.008 \\
TLI & 0.988 & & & 1.000 & & & 0.012 \\
RMSEA & 0.059 & & & 0.011 & & & -0.048 \\
\bottomrule
\end{tabular}
\label{tab:improved_model}
\end{table}

\subsection{Substantive Implications}

The baseline model suggests that fluctuations in immigration attitudes predict later changes in agreeableness, while the effect of agreeableness on immigration attitudes appears weak or unstable. This pattern is theoretically unexpected, as personality traits are generally treated as relatively stable antecedents of political attitudes.

After incorporating the 2SLW-detected residual structure, the cross-lagged estimates align more closely with theoretical expectations: within-person increases in agreeableness are associated with less exclusionary immigration attitudes, while reverse effects weaken. Thus, the 2SLW adjustment does more than improve global fit; it changes the substantive interpretation of the dynamic relationship.

Table~\ref{tab:improved_model} shows that the baseline model exhibits clear signs of misspecification, including a large $\chi^2$ statistic and a negative variance estimate, despite acceptable incremental fit indices. In contrast, the improved model yields stable and interpretable parameters and substantially improved model fit.

The latent factor $c_1$ has modest loadings, inducing a correlation slightly above 0.10. Even small residual dependencies of this magnitude can bias cross-lagged estimates in panel SEMs. The 2SLW framework provides a structured method for detecting such dependencies using formal LM and Wald tests, offering a principled alternative to ad hoc specification searches. This example illustrates how residual dependencies, even in seemingly innocuous longitudinal relationships, can bias cross-lagged estimates. The 2SLW procedure provides a structured method for detecting such bias.

\section{Discussion \& Conclusion}

Rigorous investigation of causal relationships remains central to social and behavioral research, yet unobserved confounding continues to threaten inference in panel models. Although longitudinal SEM frameworks such as the RI-CLPM provide tools for separating stable between-person differences from within-person dynamics, they do not eliminate bias arising from unmodeled dependencies. This study introduces the 2SLW approach as a structured diagnostic procedure within RI-CLPMs.

Through Monte Carlo simulations, we evaluated the performance of 2SLW under three forms of confounding: distortion of (1) predictor–outcome associations, (2) direct effects, and (3) mediation pathways. Across conditions, the procedure consistently identified model misspecification linked to unmeasured influences. An empirical application further demonstrated that incorporating 2SLW-detected residual structure can meaningfully alter substantive conclusions—not merely improve model fit. Although simulations cannot capture the full complexity of real-world data, results across both RI-CLPM and CLPM settings indicate that the approach is computationally feasible, scalable, and adaptable to a broad class of SEM-based causal panel models.

Importantly, 2SLW is a diagnostic tool rather than a confirmatory one. It identifies residual dependencies that may reflect omitted variables, measurement error, or broader structural misspecification, but it does not, on its own, establish the causal source of bias or produce bias-corrected estimates. Any model revisions prompted by 2SLW should therefore be grounded in substantive theory, supported through cross-validation, and accompanied by transparent reporting of alternative specifications. Because the first stage relies on modification indices, researchers must remain vigilant against overfitting and avoid retaining paths that capture sample-specific noise rather than theoretically meaningful structure. A natural next step is to integrate 2SLW-style diagnostics with formal sensitivity analysis frameworks that quantify how large unmeasured confounding would need to be to overturn substantive conclusions. Used together, diagnostic testing and sensitivity analysis can advance panel research toward a more explicit, systematic, and empirically disciplined evaluation of identifying assumptions. In addition, future research should examine performance in small samples, where the asymptotic properties of LM and Wald tests may produce inflated Type I error rates or insufficient power. 

\end{doublespace}

\clearpage 
\begin{singlespace}
\bibliographystyle{apacite}
\renewcommand{\refname}{}
\section*{\centering References}   
\vspace{-1cm}     
\bibliography{references}  

@article{LiuYamamoto2025,
  author    = {Licheng Liu and Teppei Yamamoto},
  title     = {Bayesian Sensitivity Analysis for Unmeasured Confounding in Causal Panel Data Models},
  journal   = {Political Analysis},
  year      = {2025},
  note      = {Advance online publication}
}

@article{bellemare2017,
  title={Lagged explanatory variables and the estimation of causal effect},
  author={Bellemare, Marc F. and Masaki, Takaaki and Pepinsky, Thomas B.},
  journal={The Journal of Politics},
  volume={79},
  number={3},
  pages={745--1171},
  year={2017},
  month={Jul},
  publisher={University of Chicago Press}
}

@article{Harring2017,
  author  = {Harring, Jeffrey R. and McNeish, Daniel M. and Hancock, Gregory R.},
  title   = {Using phantom variables in structural equation modeling to assess model sensitivity to external misspecification},
  journal = {Psychological Methods},
  year    = {2017},
  volume  = {22},
  number  = {4},
  pages   = {616--631},
  doi     = {10.1037/met0000103}
}

@article{pluemper2005,
  title={Panel data analysis in comparative politics: Linking method to theory},
  author={Pl{\"u}mper, Thomas and Troeger, Vera E. and Manow, Philip},
  journal={European Journal of Political Research},
  volume={44},
  number={2},
  pages={327--354},
  year={2005},
  publisher={Wiley},
  doi={10.1111/j.1475-6765.2005.00230.x},
  url={https://doi.org/10.1111/j.1475-6765.2005.00230.x}
}

@article{RN1041,
   author = {Acharya, Avidit and Blackwell, Matthew and Sen, Maya},
   title = {Explaining Causal Findings without Bias: Detecting and Assessing Direct Effects},
   journal = {American Political Science Review},
   volume = {110},
   pages = {512-29},
   year = {2016},
   type = {Journal Article}
}

@article{Blackwell_2014, title={A Selection Bias Approach to Sensitivity Analysis for Causal Effects}, volume={22}, DOI={10.1093/pan/mpt006}, number={2}, journal={Political Analysis}, author={Blackwell, Matthew}, year={2014}, pages={169–182}}

@article{bakker2021,
  title={Reconsidering the Link Between Self-Reported Personality Traits and Political Preferences},
  author={Bakker, Bert N. and Lelkes, Yphtach and Malka, Ariel},
  journal={American Political Science Review},
  volume={115},
  number={4},
  pages={1482--1499},
  year={2021},
  publisher={Cambridge University Press},
  doi={10.1017/S0003055421000602}
}

@inbook{RN882,
   author = {Bentler, Peter M. and Dijkstra, Theo},
   title = {Efficient Estimation via Linearization in Structural Models},
   booktitle = {Multivariate Analysis VI},
   editor = {Kirshnaiah, P. R.},
   publisher = {North-Holland},
   address = {Amsterdam},
   pages = {9-42},
   year = {1985},
   type = {Book Section}
}

@article{RN664,
   author = {Browne, Michael},
   title = {Asymptotically distribution-free methods for the analysis of covariance structures},
   journal = {British Journal of Mathematical and Statistical Psychology},
   volume = {37},
   pages = {62-83},
   DOI = {doi:10.1111/bmsp.1984.37.issue-1},
   year = {1984},
   type = {Journal Article}
}

@article{RN1108,
   author = {Dinesen, Peter Thisted and Sonderskov, Kim Mannemar and Sohlberg, Jacob and Esaiasson, Peter},
   title = {Close (Causally Connected) Cousins? Evidence on the Causal Relationship between Political trust and Social Trust },
   journal = {Public Opinion Quarterly},
   volume = {86},
   number = {3},
   pages = {708-721},
   DOI = {https://doi.org/10.1093/poq/nfac027},
   year = {2022},
   type = {Journal Article}
}

@misc{RN1101,
   author = {Feldman, Stanley and Panish, Adam and Weber, Christopher and Zheng, Bang Quan},
   title = {Rethinking the Cross Lagged Regression Model: Limitations and Alternatives in Panel Analysis},
   year = {2025},
   type = {Conference Paper}
}

@article{RN186,
   author = {Goren, Paul},
   title = {Party Identification and Core Political Values},
   journal = {American Journal of Political Science},
   volume = {49},
   number = {4},
   pages = {882-897},
   year = {2005},
   type = {Journal Article}
}

@article{RN692,
   author = {Hamaker, E. L. and Kuiper, R. M. and Grasman, R. P. P. P.},
   title = {A critique of the cross-lagged panel model},
   journal = {Psychological Methods},
   volume = {20},
   pages = {102-116},
   year = {2015},
   type = {Journal Article}
}

@article{RN1106,
   author = {Hatemi, Peter K. and Crabtree, Charles and Smith, Kevin},
   title = {Ideology Justifies Morality: Political Beliefs Predict Moral Foundations},
   journal = {American Journal of Political Science},
   volume = {63},
   number = {4},
   pages = {788-806},
   DOI = {DOI: 10.1111/ajps.12448},
   year = {2019},
   type = {Journal Article}
}

@article{RN818,
   author = {Imai, K. and Keele, L. and Tingley, D.},
   title = {A General Approach to Causal Mediation Analysis},
   journal = {Psychol Methods},
   volume = {15},
   number = {4},
   pages = {309-326},
   year = {2010},
   type = {Journal Article}
}

@article{Imai_Yamamoto_2013, title={Identification and Sensitivity Analysis for Multiple Causal Mechanisms: Revisiting Evidence from Framing Experiments}, volume={21}, DOI={10.1093/pan/mps040}, number={2}, journal={Political Analysis}, author={Imai, Kosuke and Yamamoto, Teppei}, year={2013}, pages={141–171}}

@article{RN827,
   author = {Imai, K. and Keele, L. and Tingley, D. and Yamada, T.},
   title = {Unpacking the Black Box of Causality: Learning about Causal Mechanisms from Experimental and Observational Studies},
   journal = {American Political Science Review},
   volume = {105},
   number = {4},
   pages = {765789},
   year = {2011},
   type = {Journal Article}
}

@article{RN1105,
   author = {Lucas, Richard},
   title = {Why the Cross-Lagged Panel Model Is Almost Never the Right Choice},
   journal = {Advances in Methods and Practices in Psychological Science},
   volume = {6},
   number = {1},
   pages = {1-22},
   DOI = {https://doi.org/10.1177/25152459231158},
   year = {2023},
   type = {Journal Article}
}

@article{RN1115,
   author = {Luttig, Matthew D.},
   title = {Reconsidering the Relationship between Authoritarianism and Republican Support in 2016 and Beyond},
   journal = {Journal of Politics},
   volume = {83},
   number = {2},
   pages = {783-787},
   year = {2021},
   type = {Journal Article}
}

@article{RN1109,
   author = {Mund, M. and Johnson, M. D. and Nestler, S. },
   title = {Changes in Size and Interpretation of Parameter Estimates in Within-Person Models in the Presence of Time-Invariance and Time-Varying Covariate},
   journal = {Frontiers in Psychology},
   DOI = {https://doi.org/https://doi.org/10.3389/fpsyg.2021.666928 },
   year = {2021},
   type = {Journal Article}
}

@article{RN547,
   author = {Rosseel, Yves},
   title = {lavaan: An R package for structural equation modeling},
   journal = {Journal of Statistical Software},
   volume = {48},
   number = {2},
   pages = {1-36},
   year = {2012},
   type = {Journal Article}
}

@article{RN689,
   author = {Usami, Satoshi and Murayama, K and Hamaker, E. L. },
   title = {A unified framework of longitudinal models to examine reciprocal relations},
   journal = {Psychol Methods},
   volume = {24},
   number = {5},
   pages = {637-657},
   year = {2019},
   type = {Journal Article}
}

@article{RN708,
   author = {Zheng, Bang Quan and Bentler, Peter M.},
   title = {Testing Mean and Covariance Structures with Reweighted Least Squares},
   journal = {Structural Equation Modeling: A Multidisplinary Journal},
   DOI = {https://doi.org/10.1080/10705511.2021.1977649},
   year = {2021},
   type = {Journal Article}
}

@article{RN949,
   author = {Zheng, Bang Quan and Bentler, Peter M.},
   title = {{RGLS} and {RLS} in Covariance Structure Analysis},
   journal = {Structural Equation Modeling: A Multidisplinary Journal},
   volume = {30},
   number = {2},
   pages = {234-244},
   DOI = {https://doi.org/10.1080/10705511.2022.2117182},
   url = {SocArXiv https://doi.org/10.31235/osf.io/aejgf},
   year = {2023},
   type = {Journal Article}
}

@article{RN1095,
   author = {Zheng, Bang Quan and Bentler, Peter M.},
   title = {Enhancing Model Fit Evaluation in SEM: Practical Tips for Optimizing Chi-Square Tests},
   journal = {Structural Equation Modeling: A Multidisciplinary Journal},
   number = {Advance online publication},
   DOI = {https://www.tandfonline.com/doi/full/10.1080/10705511.2024.2354802},
   year = {2024},
   type = {Journal Article}
}
\end{singlespace}

\clearpage

\section*{\centering{ Appendix}}

\renewcommand{\thetable}{A\arabic{table}}
\setcounter{table}{0}

\renewcommand{\thefigure}{A\arabic{figure}}
\setcounter{figure}{0} 

\subsection*{The Covariance Matrix of the RI-CLPM}

In covariance structure analysis, the RI-CLPM can be expressed through its covariance matrix. Specifically, Eqs. (5) and (6) can be reformulated in matrix form as follows:

\begin{align}
z_{it} &= \delta_i + \Phi z_{i(t-1)}^* + \eta_i + v_{it}, & (i=1,\dots,n, \ t=1,\dots,T) \notag \\
z_{i1} &= \delta_i + \Psi \eta_i + \epsilon_{i1}, & (i=1,\dots,n) \tag{A1}
\end{align}

where

\begin{align}
z_{it} &= \begin{bmatrix} x_{it} \\ y_{it} \end{bmatrix}, &
\Phi &= \begin{bmatrix} \beta_x & \gamma_x \\ \beta_y & \gamma_y \end{bmatrix}, &
\eta_i &= \begin{bmatrix} \eta_{xi} \\ \eta_{yi} \end{bmatrix}, &
v_i &= \begin{bmatrix} v_{xit} \\ v_{yit} \end{bmatrix}, &
\delta_i &= \begin{bmatrix} \delta_{xi} \\ \delta_{yi} \end{bmatrix}, \tag{A2} \\
\Psi &= \begin{bmatrix} \psi_{11} & \psi_{12} \\ \psi_{21} & \psi_{22} \end{bmatrix}, &
\epsilon_{i0} &= \begin{bmatrix} \epsilon_{i0}^y \\ \epsilon_{i0}^x \end{bmatrix}. \tag{A3}
\end{align}

Building on the CLPM covariance matrix formulation in Hayakawa (2019), we assume that the largest eigenvalue of $\Phi$ is less than 1, ensuring that $z_{it}$ follows a stable process. Using these matrices, we combine Eqs. (5) and (6) to obtain:

\begin{equation}
B z_i = \delta_i + J \pi_i + u_i, \tag{A4}
\end{equation}

where 

\begin{equation}
B = 
\begin{bmatrix}
I_2 & 0 & \cdots & 0 \\
-\Phi & I_2 & \cdots & \vdots \\
\vdots & \ddots & \ddots & 0 \\
0 & 0 & -\Phi & I_2
\end{bmatrix} \tag{A5}
\end{equation}

\vspace{0.8cm}

\begin{equation}
\Gamma = B^{-1} = 
\begin{bmatrix}
I_2 & 0 & \cdots & 0 \\
\Phi & I_2 & \cdots & \vdots \\
\vdots & \ddots & \ddots & 0 \\
\Phi^{T-1} & 0 & \Phi & I_2
\end{bmatrix} \tag{A6}
\end{equation}

\vspace{0.8cm}

\begin{equation}
z_i =
\begin{bmatrix} z_{i1} \\ z_{i2} \\ \vdots \\ z_{iT} \end{bmatrix}, \quad
u_i = 
\begin{bmatrix} \epsilon_{i1} \\ v_{i2} \\ \vdots \\ v_{iT} \end{bmatrix} \tag{A7}
\end{equation}

\vspace{0.8cm}

\begin{equation}
J =
\begin{bmatrix} 1 & \Psi \\ 0 & I_2 \end{bmatrix}, \quad
\pi_i = 
\begin{bmatrix} \delta_i \\ \eta_i \end{bmatrix} \tag{A8}
\end{equation}

Thus, we have:

\begin{equation}
z_i = \Gamma (J \pi_i + u_i). \tag{A9}
\end{equation}

We assume that $\pi_i \sim (0, \Sigma_\pi)$, $v_{it} \sim (0, \Sigma_v)$, $\epsilon_{i0} \sim (0, \Sigma_\epsilon)$, and that $\delta_i$, $\eta_i$, $v_{it}$, and $\epsilon_{i1}$ are uncorrelated, independent, and identically distributed. Therefore, the covariance matrix can be written as:

\begin{equation}
\Sigma(\theta) = \Gamma \left( J \Sigma_\pi J' + \Sigma_u \right) \Gamma', \tag{A10}
\end{equation}

where 
\[
\theta = (\Phi, \Psi, \sigma_\pi^2, \sigma_v^2, \sigma_\epsilon^2)', \quad 
\Sigma_u = \mathrm{diag}(\sigma_\epsilon^2, \sigma_v^2 I_T).
\]

\section*{Monte Carlo Simulation}

The data generation scheme for the simulation follows the covariance matrix of the RI-CLPM described earlier. Specifically, the data are generated as
\[
z_i = \Sigma(\theta)^{1/2} \, \varepsilon_i,
\]
where $\Sigma(\theta)$ is the covariance matrix and $\varepsilon_i$ follows a standard normal distribution, $\varepsilon_i \sim N(0,1)$. 

For the parameter values, the autoregressive (AR) parameters $\beta_{xi}$ and cross-lagged (CL) parameters $\gamma_{yi}$ are set to 0.25 and 0.15, respectively. We use identical parameter values and specifications for the time-varying component of the RI-CLPM across all models to ensure comparability and to validate the simulation results. 

Using identical parameter values in the simulation facilitates validation and allows for comparison with existing studies. While the choice of parameter values in simulations can be arbitrary, it does not affect test statistics or fit indices when models are correctly specified. Data generation can be implemented in \texttt{R} using the \texttt{lavaan} package.

\section*{Five-Wave RI-CLPM with Two Indicators Per Factor}

To assess the flexibility of the 2SLW approach, we construct a more complex RI-CLPM featuring five waves of repeated measurements, where each latent factor is measured by two indicators. Additionally, we specify a second-order factor $L$, which is measured by the factors $\eta_x$ and $\eta_y$. 

Following a similar strategy introduced in the main text, we design three scenarios: 
\begin{enumerate}
    \item The correlation between the two factors is driven by an unobserved latent confounder.
    \item The association is modeled as a direct effect between the two factors.
    \item The relationship follows a mediation path between two time-varying factors.
\end{enumerate}

Consider two constructs, $X$ and $Y$, measured at $t=1,2$.  
Each observed score is represented as

\begin{equation}
\begin{aligned}
x_{it} &= \delta_{xt} + \eta_{xi} + x_{it}^*, \\
y_{it} &= \delta_{yt} + \eta_{yi} + y_{it}^*,
\end{aligned} \tag{A11}
\end{equation}

with $x_{it}^* \perp y_{is}^*$.

In the cross-lagged panel model (CLPM), the within-person deviations evolve according to

\begin{equation}
\begin{aligned}
x_{it}^* &= \beta_x x_{i(t-1)}^* + \gamma_x y_{i(t-1)}^* + v_{xit}, \\
y_{it}^* &= \beta_y y_{i(t-1)}^* + \gamma_y x_{i(t-1)}^* + v_{yit},
\end{aligned} \tag{A12}
\end{equation}

where the disturbances $v_{xit}, v_{yit}$ have mean zero and are independent of measurement errors.

In the random-intercept CLPM (RI-CLPM), $\eta_{xi}$ and $\eta_{yi}$ are treated as time-invariant person-specific components (random intercepts), while the within-person deviations $x_{it}^*, y_{it}^*$ follow the same autoregressive and cross-lagged dynamics:

\begin{equation}
\begin{aligned}
x_{it}^* &= \beta_x x_{i(t-1)}^* + \gamma_x y_{i(t-1)}^* + v_{xit}, \\
y_{it}^* &= \beta_y y_{i(t-1)}^* + \gamma_y x_{i(t-1)}^* + v_{yit}.
\end{aligned} \tag{A13}
\end{equation}

Let $\mathbf{\Sigma}(\theta)$ denote the model-implied covariance matrix of the observed vector $z_i = (x_{i1}, y_{i1}, x_{i2}, y_{i2})^\top$, expressed as a smooth function of the parameter vector  

\[
\theta = (\beta_x, \beta_y, \gamma_x, \gamma_y, \ldots).
\]

Under multivariate normality, the expected Fisher information for any structural parameter $\psi \in \{\beta_x, \beta_y, \gamma_x, \gamma_y\}$ is  

\begin{equation}
I(\psi) = \tfrac{1}{2}\,\mathrm{tr}\!\left(\mathbf{\Sigma}^{-1} \frac{\partial \mathbf{\Sigma}}{\partial \psi}\,\mathbf{\Sigma}^{-1} \frac{\partial \mathbf{\Sigma}}{\partial \psi}\right). \tag{A14}
\end{equation}

Identification requires that the Jacobian  

\begin{equation}
J(\theta) = \big[\, \mathrm{vec}\!\left(\tfrac{\partial \mathbf{\Sigma}}{\partial \theta_j}\right) : \theta_j \in \theta \,\big] \tag{A15}
\end{equation}

has full column rank.

\subsection*{Proposition 1 (Weak loadings $\lambda_x \to 0$ and/or $\lambda_y \to 0$)}

Consider single-indicator CLPM or RI-CLPM, where
\[
x_{it} = \delta_{xt} + \eta_{xi} + \lambda_x x_{it}^*, \quad
y_{it} = \delta_{yt} + \eta_{yi} + \lambda_y y_{it}^*,
\]
with within-person deviations $x_{it}^*, y_{it}^*$ evolving according to the cross-lag dynamics.  

For any structural parameter $\psi \in \{\beta_x, \beta_y, \gamma_x, \gamma_y\}$, the derivative of the model-implied covariance matrix scales entrywise with the loadings:
\begin{equation}
\frac{\partial \mathbf{\Sigma}}{\partial \psi} = O(\lambda_x \lambda_y). \tag{A17}
\end{equation}

Hence, the expected Fisher information satisfies
\begin{equation}
I(\psi) = O(\lambda_x^2 \lambda_y^2) \to 0 \tag{A18}
\end{equation}
as any loading approaches zero, implying near-zero information and near-nonidentification of $\psi$.

\paragraph{Sketch of proof.}  
Observed covariances inherit structural dependencies via the measurement equation:
\[
\mathrm{cov}(x_s, y_t) = \lambda_x \lambda_y \, \mathrm{cov}(x_s^*, y_t^*), \tag{A19}
\]
so differentiating w.r.t.\ $\psi$ introduces the $\lambda_x \lambda_y$ factor.  
Thus $I(\psi) \sim (\lambda_x \lambda_y)^2$. \hfill $\blacksquare$

\subsection*{Proposition 2 (Extremely strong loadings $\lambda_x, \lambda_y \to 1$)}

In single-indicator CLPM/RI-CLPM, when $\lambda_x = \lambda_y \approx 1$ and residuals $v_{xit}, v_{yit} \approx 0$, the columns of $\mathbf{J}(\theta)$ corresponding to measurement and structural parameters become nearly collinear, making the Jacobian ill-conditioned.  
In RI-CLPM, this includes collinearity between random-intercept variance components and within-person dynamics, yielding weak identification of $(\beta_x, \beta_y, \gamma_x, \gamma_y)$.

\paragraph{Sketch of proof (CLPM).}  
With $\lambda \approx 1$ and negligible measurement error, $x_{it} \approx x_{it}^*, \, y_{it} \approx y_{it}^*$.  
Changes in $\mathbf{\Sigma}$ from adjusting a structural parameter $\psi$ are nearly indistinguishable from changes caused by latent variance parameters.  
Thus columns of $\mathbf{J}(\theta)$ are nearly linearly dependent.

\paragraph{Sketch of proof (RI-CLPM).}  
Observed scores are
\begin{equation}
\begin{aligned}
x_{it} &= \delta_{xt} + \eta_{xi} + \lambda_x x_{it}^*, \\
y_{it} &= \delta_{yt} + \eta_{yi} + \lambda_y y_{it}^*,
\end{aligned} 
\tag{A20}
\end{equation}
with $\lambda_x, \lambda_y \approx 1$ and $v_{xit}, v_{yit} \approx 0$.  
Perturbations in $\mathrm{Var}(\eta_{xi})$ or lag parameters $(\beta_x, \gamma_x)$ alter almost the same off-diagonal elements of $\mathbf{\Sigma}$, yielding an ill-conditioned Jacobian, small Fisher information, and unstable estimates. \hfill $\blacksquare$

\subsection*{Corollary (Extremes $\Rightarrow$ instability)}

Under either extreme (Proposition 1 or 2), the Fisher information for cross-lag parameters vanishes or the Jacobian is ill-conditioned.  
Therefore MLEs (or GLS/WLS estimators) of $(\beta_x, \beta_y, \gamma_x, \gamma_y)$ have exploding variance and are highly sensitive to small data perturbations.  
In practice, this manifests as:
\begin{itemize}
\item Cross-lag coefficients that flip sign across specifications or samples;
\item Path diagrams whose latent-observed linkages appear “random”;
\item Convergence to different local optima depending on starting values.
\end{itemize}

\subsection*{Practical Identification Conditions}

To avoid these degeneracies in CLPM/RI-CLPM:
\begin{itemize}
\item Within-person deviations $x_{it}^*, y_{it}^*$ of moderate size with nontrivial residuals $v_{xit}, v_{yit}$;
\item Multiple indicators per construct per wave (providing additional degrees of freedom);
\item Equality constraints on measurement components across waves to fix scale;
\item In RI-CLPM, sufficient time points and variation to separately identify between- and within-person components.
\end{itemize}

These conditions ensure $\mathbf{J}(\theta)$ has full rank and good conditioning, so Fisher information is well-behaved and estimates are stable.

\newpage

\subsection*{Parameter Estimates based on Table 2, 3, and 4}

\begin{table}[htbp]
\centering
\caption{Five-Wave RI-CLPM: AR \& CL Effects}
\begin{tabular}{lccc}
\hline
 & Estimate & Std. Err & $P$-value \\
\hline
\multicolumn{4}{l}{WFX$_2$ $\sim$} \\
\quad WFX$_1$ & 0.196 & 0.060 & 0.001 \\
\quad WFY$_1$ & 0.192 & 0.056 & 0.000 \\

\multicolumn{4}{l}{WFY$_2$ $\sim$} \\
\quad WFY$_1$ & 0.246 & 0.059 & 0.000 \\
\quad WFX$_1$ & 0.192 & 0.050 & 0.000 \\

\multicolumn{4}{l}{WFX$_3$ $\sim$} \\
\quad WFX$_2$ & 0.235 & 0.052 & 0.000 \\
\quad WFY$_2$ & 0.228 & 0.048 & 0.000 \\

\multicolumn{4}{l}{WFY$_3$ $\sim$} \\
\quad WFY$_2$ & 0.316 & 0.051 & 0.000 \\
\quad WFX$_2$ & 0.159 & 0.045 & 0.000 \\

\multicolumn{4}{l}{WFX$_4$ $\sim$} \\
\quad WFX$_3$ & 0.204 & 0.041 & 0.000 \\
\quad WFY$_3$ & 0.229 & 0.040 & 0.000 \\

\multicolumn{4}{l}{WFY$_4$ $\sim$} \\
\quad WFX$_3$ & 0.150 & 0.120 & 0.000 \\
\quad WFY$_3$ & 0.126 & 0.125 & 0.002 \\

\multicolumn{4}{l}{WFX$_4$} \\
\quad WFX$_2$ & 0.497 & 0.035 & 0.000 \\

\hline
\midrule
\multicolumn{4}{l}{Model Fit} \\
\midrule
$\chi^2$   & 12.954 & & \\
$\mathit{df}$         & 8      & & \\
CFI        & 0.998  & & \\
NFI        & 0.995  & & \\
TLI        & 0.994  & & \\
RMSEA      & 0.025  & & \\
\hline
\bottomrule
\end{tabular}
\end{table}

\begin{table}[htbp]
\centering
\caption{Five-Wave RI-CLPM: Mediation Paths}
\begin{tabular}{lccc}
\hline
 & Estimate & Std. Err & $P$-value \\
\hline
\multicolumn{4}{l}{WFX$_2$ $\sim$} \\
\quad WFX$_1$ & 0.280 & 0.051 & 0.000 \\
\quad WFY$_1$ & 0.216 & 0.050 & 0.000 \\

\multicolumn{4}{l}{WFY$_2$ $\sim$} \\
\quad WFY$_1$ & 0.216 & 0.063 & 0.000 \\
\quad WFX$_1$ & 0.279 & 0.049 & 0.000 \\

\multicolumn{4}{l}{WFX$_3$ $\sim$} \\
\quad WFX$_2$ & 0.296 & 0.048 & 0.000 \\
\quad WFY$_2$ & 0.228 & 0.045 & 0.000 \\

\multicolumn{4}{l}{WFY$_3$ $\sim$} \\
\quad WFY$_2$ & 0.249 & 0.054 & 0.000 \\
\quad WFX$_2$ & 0.274 & 0.047 & 0.000 \\

\multicolumn{4}{l}{WFX$_4$ $\sim$} \\
\quad WFX$_3$ & 0.336 & 0.041 & 0.000 \\
\quad WFY$_3$ & 0.268 & 0.040 & 0.000 \\

\multicolumn{4}{l}{WFY$_4$ $\sim$} \\
\quad WFY$_3$ & -0.067 & 0.435 & 0.642 \\
\quad WFX$_3$ & 0.514 & 0.860 & 0.074 \\
\quad WFY$_2$ & 0.200 & 0.212 & 0.005 \\

\hline
\midrule
\multicolumn{4}{l}{Model Fit} \\
\midrule
$\chi^2$   & 6.723 & & \\
$\mathit{df}$         & 7     & & \\
CFI        & 1.00  & & \\
NFI        & 0.997 & & \\
TLI        & 1.00  & & \\
RMSEA      & 0.00  & & \\
\hline
\bottomrule
\end{tabular}
\end{table}

\begin{table}[htbp]
\centering
\caption{Five-Wave RI-CLPM: Covariance Structure}
\begin{tabular}{lccc}
\hline
 & Estimate & Std. Err & $P$-value \\
\hline
\multicolumn{4}{l}{WFX$_2$ $\sim$} \\
\quad WFX$_1$ & 0.125 & 0.071 & 0.000 \\
\quad WFY$_1$ & 0.115 & 0.065 & 0.000 \\

\multicolumn{4}{l}{WFY$_2$ $\sim$} \\
\quad WFY$_1$ & 0.091 & 0.070 & 0.004 \\
\quad WFX$_1$ & 0.165 & 0.070 & 0.000 \\

\multicolumn{4}{l}{WFX$_3$ $\sim$} \\
\quad WFX$_2$ & 0.472 & 0.015 & 0.000 \\
\quad WFY$_2$ & 0.407 & 0.015 & 0.000 \\

\multicolumn{4}{l}{WFY$_3$ $\sim$} \\
\quad WFY$_2$ & 0.411 & 0.015 & 0.000 \\
\quad WFX$_2$ & 0.482 & 0.014 & 0.000 \\

\multicolumn{4}{l}{WFX$_4$ $\sim$} \\
\quad WFX$_3$ & 0.125 & 0.066 & 0.000 \\
\quad WFY$_3$ & 0.192 & 0.068 & 0.000 \\

\multicolumn{4}{l}{WFY$_4$ $\sim$} \\
\quad WFY$_3$ & 0.172 & 0.065 & 0.000 \\
\quad WFX$_3$ & 0.138 & 0.060 & 0.000 \\

\multicolumn{4}{l}{Covariances} \\
\quad WFX$_4$ ~~ WFY$_2$ & 0.803 & 0.022 & 0.000 \\
\quad WFX$_2$ ~~ WFY$_4$ & 0.810 & 0.023 & 0.000 \\

\hline
\midrule
\multicolumn{4}{l}{Model Fit} \\
$\chi^2$   & 5.753 & & \\
$\mathit{df}$         & 7     & & \\
CFI        & 1.00  & & \\
NFI        & 0.999 & & \\
TLI        & 1.00  & & \\
RMSEA      & 0.00  & & \\
\hline
\bottomrule
\end{tabular}
\end{table}

\subsection*{5-Wave RI-CLPM with 2 Indicators Per Construct}

In this section, we test the 2SLW approach on the higher-order RI-CLPM, that is, we create an latent variable based on the between-person components. Prior to testing the three scenarios, we create a baseline model, which is based on Figure~A1 but without the dashed lines. As emphasized by previous studies, SEM-based panel models, including RI-CLPM, CLPM, latent growth models, etc., are based on parametric and asymptotic properties. That is, they are linear and follow a distribution. Asymptotic properties refer to how an estimator or test behaves as the sample size approaches infinity, with key features such as consistency—where the estimator converges in probability to the true parameter—providing theoretical justification for inference when finite-sample behavior is uncertain or complex.

\begin{figure}[H]
    \centering
    \includegraphics[width=0.7\textwidth]{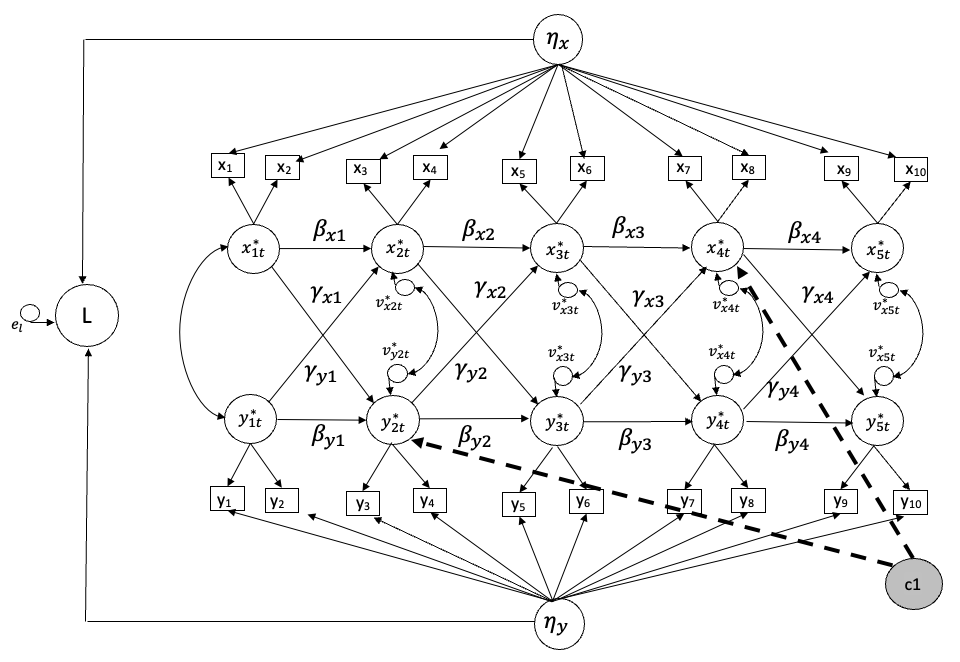} 
    \caption{5-Wave RI-CLPM with 2 Indicators (Correlation)}
    \label{fig:A1}
\end{figure}


\renewcommand{\arraystretch}{0.75}
\begin{table}[H]
\centering
\caption{Monte Carlo and Asymptotic Results for 5-Wave RI-CLPM}
\begin{tabular}{cccccccc}  
\toprule
N & {LM $\chi^2$} & SD & $P$-value & Rej Rate & NFI & CFI & RMSEA \\
\midrule
100    & 186.332 & 19.895 & 0.273 & 0.215 & 0.841 & 0.993 & 0.020 \\
200    & 177.391 & 18.738 & 0.389 & 0.101 & 0.922 & 1.000 & 0.000 \\
300    & 175.550 & 18.554 & 0.415 & 0.078 & 0.945 & 1.000 & 0.000 \\
500    & 172.176 & 19.222 & 0.472 & 0.070 & 0.960 & 0.997 & 0.011 \\
800    & 172.169 & 19.271 & 0.469 & 0.068 & 0.974 & 0.997 & 0.012 \\
1000   & 171.764 & 17.909 & 0.476 & 0.066 & 0.982 & 1.000 & 0.000 \\
5000   & 171.068 & 18.024 & 0.490 & 0.062 & 0.996 & 1.000 & 0.003 \\
10000  & 169.742 & 18.433 & 0.505 & 0.050 & 0.998 & 1.000 & 0.000 \\
\hline
\bottomrule
\end{tabular}
\label{tab:table_a1}
\caption*{\footnotesize Note: Theoretically expected mean $\chi^2 = 170$, SD = 18.44, $P$-value = 0.5, rejection rate = 0.05.}
\end{table}


\renewcommand{\arraystretch}{0.75}
\begin{table}[H]
\centering
\caption{Test Result for 5-Wave RI-CLPM with 2 Indicators (Correlation)}
\vspace{-0.3cm}
\begin{tabular}{lllllll}
\toprule
Parameter & Operator & Predictor & LM & EPC & Wald Test & $P$-value \\
\midrule
\rowcolor{gray!20} WFX$_4$ & $\sim\sim$ & WFY$_2$ & 42.594 & 0.468 & 9.597 & 0.002 \\
WFX$_5$ &  $\sim$ & WFY$_2$ & 33.129 & -0.337 & 1.839 & 0.175 \\
WFX$_5$ &  $\sim\sim$ & WFY$_2$ & 29.012 & -0.374 & 0.676 & 0.411 \\
WFX$_4$ &  $\sim$ & WFX$_1$ & 12.763 & -0.268 & 0.111 & 0.739 \\
WFX$_4$ & $=\sim$ & Y$_{22}$ & 8.778 & 0.141 & 0.616 & 0.433 \\
WFX$_1$ &  $\sim\sim$ & WFX$_4$ & 8.008 & -0.236 & 0.010 & 0.920 \\
WFY$_3$ &  $=\sim$ & X$_{32}$ & 6.658 & 0.155 & 3.516 & 0.061 \\
X$_{31}$ &  $\sim\sim$ & Y$_{32}$ & 6.190 & -0.122 & 2.493 & 0.114 \\
\hline
\bottomrule
\end{tabular}
\label{tab:table_a2}
\end{table}

Figure~\ref{fig:A1} illustrates a hypothetical scenario where a latent confounder, $c_1$, affects the residuals of WFX$_4$ and WFY$_2$. Using this specification, we generate a population model. we then estimate an analysis model that omits $c_1$ and its direct effects on the residuals of WFX$_4$ and WFY$_2$. 

If the 2SLW approach functions as intended, it should detect the residual correlation between WFX$_4$ and WFY$_2$ induced by the omitted confounder. However, this relationship is not always straightforward. In some cases, the LM test may identify a correlation or directional effect between WFX$_4$ and WFY$_2$ that appears statistically significant, even when the underlying causal structure is more complex. 

The final selection between these specifications can be guided by overall model fit, comparing a model that includes a direct effect to one that specifies a correlation between the factors. Naturally, including the true confounder(s) in the model will improve the $\chi^2$ test statistic and overall fit indices.

Table~\ref{tab:table_a1} shows that the average of the $\chi^2$ test statistics approaches the expected value of 170 when the sample size is sufficiently large. The standard deviation of the $\chi^2$ test statistics, the average $p$-value, and the rejection rate are also close to their expected values of 18.44, 0.5, and 0.05, respectively. In addition, the CFI, NFI, and TLI are close to 1, while the RMSEA is near 0. Taken together, these results suggest that the model is well specified.

Table~\ref{tab:table_a2} shows the 2SLW results, as we can see that it successfully identified the correlation between WFX$_4$ and WFY$_2$.

\renewcommand{\arraystretch}{0.75}
\begin{table}[H]
\centering
\caption{AR \& CL Effects – 5-Wave RI-CLPM (Correlation)}
\vspace{-0.3cm}
\small 
\begin{tabular}{lccc}
\toprule
Parameter & Estimate & Std. Err & $P$-value \\
\midrule
WFX$_2$ $\sim$ & & & \\
WFX$_1$ & 0.354 & 0.063 & 0.000 \\
WFY$_1$ & 0.123 & 0.062 & 0.026 \\
WFY$_2$ $\sim$ & & & \\
WFY$_1$ & 0.175 & 0.067 & 0.003 \\
WFX$_1$ & 0.017 & 0.066 & 0.788 \\
WFX$_3$ $\sim$ & & & \\
WFX$_2$ & 0.253 & 0.054 & 0.000 \\
WFY$_2$ & 0.152 & 0.054 & 0.022 \\
WFY$_3$ $\sim$ & & & \\
WFY$_2$ & 0.239 & 0.053 & 0.000 \\
WFX$_2$ & 0.249 & 0.052 & 0.000 \\
WFX$_4$ $\sim$ & & & \\
WFX$_3$ & 0.003 & 0.107 & 0.975 \\
WFY$_3$ & 0.106 & 0.076 & 0.107 \\
WFY$_4$ $\sim$ & & & \\
WFY$_3$ & 0.143 & 0.072 & 0.054 \\
WFX$_3$ & 0.195 & 0.084 & 0.013 \\
WFX$_5$ $\sim$ & & & \\
WFX$_4$ & 0.225 & 0.056 & 0.001 \\
WFY$_4$ & 0.162 & 0.072 & 0.020 \\
WFY$_5$ $\sim$ & & & \\
WFY$_4$ & 0.140 & 0.074 & 0.084 \\
WFX$_4$ & 0.273 & 0.050 & 0.000 \\
\hline
\midrule
\multicolumn{4}{l}{Model Fit} \\
\midrule
$\chi^2$ & 155.733 & & \\
$\mathit{df}$  & 169 & & \\
CFI & 1.000 & & \\
NFI & 0.983 & & \\
TLI & 1.002 & & \\
RMSEA & 0.000 & & \\
\hline
\bottomrule
\end{tabular}
\label{tab:table_a3}
\end{table}

\begin{figure}[H]
    \centering
    \caption{5-Wave RI-CLPM with 2 Indicators (Direct Effect)}
    \includegraphics[width=0.7\textwidth]{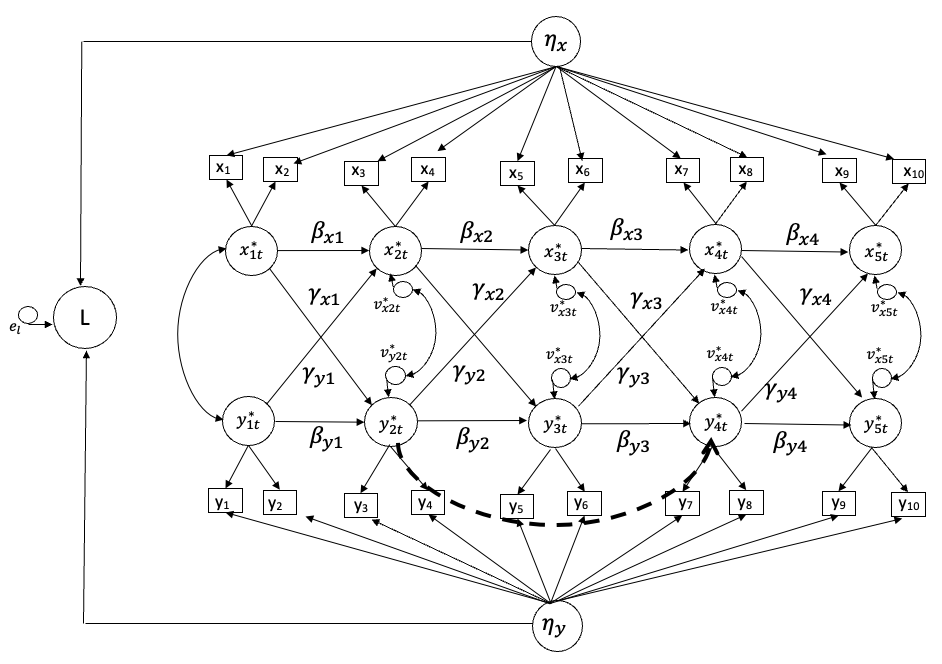} 
    \label{fig:A2}
\end{figure}

\begin{table}[H]
\centering
\caption{Test Result for 5-Wave RI-CLPM with 2 Indicators (Direct Effect)}
\vspace{-0.3cm}
\begin{tabular}{lllllll}
\toprule
Parameter & Operator & Predictor & LM $\chi^2$ & EPC & Wald Test & $P$-value \\
\midrule
\rowcolor{gray!20} WFX$_5$ &  $\sim$ & WFX$_2$ & 113.635 & 1.022 & 209.231 & 0.000 \\
WFX$_2$ &  $\sim\sim$ & WFX$_5$ & 95.532 & 0.690 & 2.881 & 0.090 \\
WFX$_4$ &  $\sim$ & WFX$_2$ & 94.187 & -0.827 & 0.970 & 0.325 \\
\rowcolor{gray!20} WFX$_2$ &  $\sim\sim$ & WFX$_4$ & 79.406 & -0.554 & 11.150 & 0.001 \\
WFY$_4$ &  $\sim$ & WFY$_2$ & 29.105 & 0.867 & 0.413 & 0.520 \\
WFY$_2$ &  $\sim\sim$ & WFY$_4$ & 25.333 & 0.441 & 0.413 & 0.520 \\
\hline
\bottomrule
\end{tabular}
\label{tab:table_a4}
\end{table}

\renewcommand{\arraystretch}{0.75}
\begin{table}[H]
\centering
\caption{AR \& CL Effects – 5-Wave RI-CLPM (Direct Effect)}
\vspace{-0.3cm}
\begin{tabular}{lccc}
\toprule
Parameter & Estimate & Std. Err & $P$-value \\
\midrule
WFX$_2$ $\sim$ & & & \\
WFX$_1$ & 0.407 & 0.062 & 0.000 \\
WFY$_1$ & 0.226 & 0.069 & 0.000 \\
WFY$_2$ $\sim$ & & & \\
WFY$_1$ & 0.155 & 0.069 & 0.037 \\
WFX$_1$ & 0.281 & 0.059 & 0.000 \\
WFX$_3$ $\sim$ & & & \\
WFX$_2$ & 0.262 & 0.070 & 0.001 \\
WFY$_2$ & 0.264 & 0.080 & 0.000 \\
WFY$_3$ $\sim$ & & & \\
WFY$_2$ & 0.126 & 0.083 & 0.134 \\
WFX$_2$ & 0.092 & 0.055 & 0.201 \\
WFX$_4$ $\sim$ & & & \\
WFX$_3$ & 0.272 & 0.118 & 0.057 \\
WFY$_3$ & -0.003 & 0.097 & 0.980 \\
WFY$_4$ $\sim$ & & & \\
WFY$_3$ & 0.118 & 0.069 & 0.038 \\
WFX$_3$ & 0.200 & 0.062 & 0.001 \\
WFX$_5$ $\sim$ & & & \\
WFX$_4$ & 0.131 & 0.077 & 0.011 \\
WFY$_4$ & 0.177 & 0.052 & 0.000 \\
WFY$_5$ $\sim$ & & & \\
WFY$_4$ & 0.184 & 0.063 & 0.021 \\
WFX$_4$ & -0.020 & 0.115 & 0.860 \\
WFX$_5$ $\sim$ & & & \\
WFX$_2$ & 0.625 & 0.047 & 0.000 \\
WFY$_4$ $\sim$ & & & \\
WFY$_2$ & 0.423 & 0.078 & 0.000 \\
\hline
\midrule
\multicolumn{4}{l}{Model Fit} \\
\midrule
$\chi^2$ & 167.117 & & \\
$\mathit{df}$  & 168 & & \\
CFI & 1.000 & & \\
NFI & 0.983 & & \\
TLI & 1.000 & & \\
RMSEA & 0.000 & & \\
\hline
\bottomrule
\end{tabular}
\caption*{\footnotesize Note: All coefficients are standardized.}
\label{tab:table_a5}
\end{table}

\clearpage 

\section*{Five-Wave RI-CLPM with Two Indicators Per Factor (Mediation)}

Figure~\ref{tab:table_a6} illustrates that the 2SLW approach can identify the mediation path in the five-wave RI-CLPM with two indicators per factor. In this model, $M$ functions as the mediator between the residuals of WFX$_4$ and WFX$_2$. 

Table~A6 presents the 2SLW results, showing that the previously unmeasured mediation path is identified. Moreover, the mediator is regressed on $x_1$, revealing a statistically significant relationship between WFX$_4$ and WFX$_2$. 

To further validate this parameter, it should be incorporated into the analysis model to assess both its statistical significance and substantive relevance.

\begin{figure}[H]
    \centering
    \caption{5-Wave RI-CLPM with 2 Indicators (Mediation)}
    \includegraphics[width=0.7\textwidth]{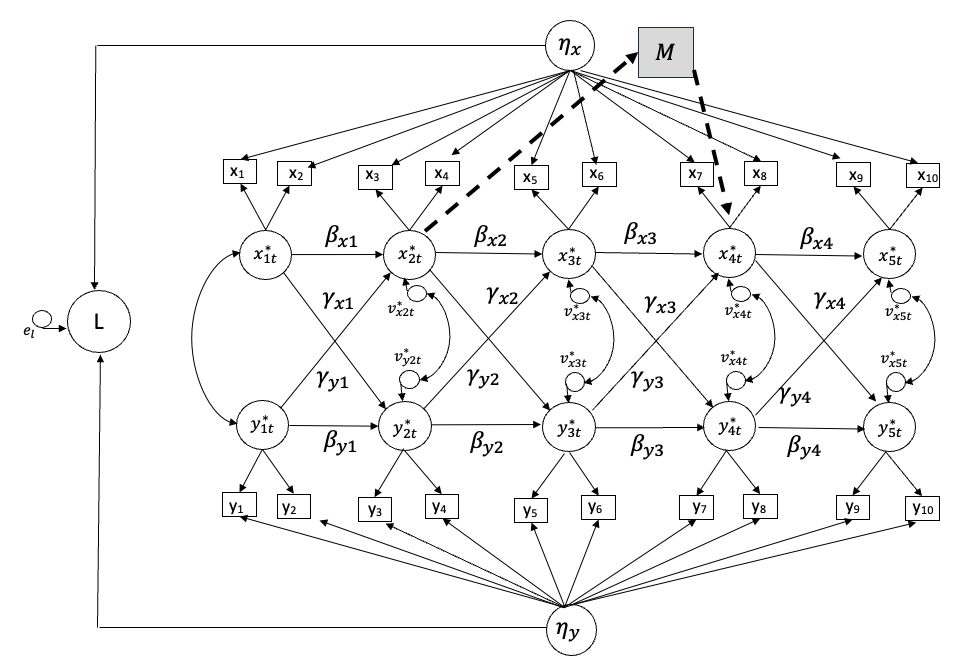} 
    \label{fig:}
    \label{fig:figure_a3}
\end{figure}

\renewcommand{\arraystretch}{0.75}
\begin{table}[H]
\centering
\caption{Test Result for 5-Wave RI-CLPM with 2 Indicators (Mediation)}
\vspace{-0.3cm}
\begin{tabular}{lllllll}
\toprule
Parameter & Operator & Predictor & LM $\chi^2$ & EPC & Wald Test & $P$-value \\
\midrule
\rowcolor{gray!20} WFX$_4$ &  $\sim$ & WFX$_2$ & 24.742 & 0.747 & 72.181 & 0.000 \\
WFX$_2$ &  $\sim\sim$ & WFX$_5$ & 22.761 & -0.302 & 2.520 & 0.112 \\
WFX$_5$ &  $\sim$ & WFX$_2$ & 21.493 & -0.461 & 0.075 & 0.784 \\
\rowcolor{gray!20} WFX$_1$ &  $\sim\sim$ & WFX$_4$ & 12.717 & 0.248 & 61.918 & 0.000 \\
x$_{31}$  &  $\sim\sim$ & x$_{32}$  & 9.273  & 5.633 & 0.017 & 0.898 \\
\hline
\bottomrule
\end{tabular}
\label{tab:table_a6}
\end{table}

\renewcommand{\arraystretch}{0.75}
\begin{table}[htbp]
\centering
\caption{AR \& CL Effects – 5-Wave RI-CLPM (Mediation)}
\vspace{-0.3cm}
\begin{tabular}{lccc}
\toprule
Parameter & Estimate & Std. Err & $P$-value \\
\midrule
WFX$_2$ $\sim$ & & & \\
\quad WFX$_1$ & 0.085 & 0.081 & 0.275 \\
\quad WFY$_1$ & 0.187 & 0.072 & 0.010 \\
WFY$_2$ $\sim$ & & & \\
\quad WFY$_1$ & 0.248 & 0.072 & 0.000 \\
\quad WFX$_1$ & 0.065 & 0.072 & 0.325 \\
WFX$_3$ $\sim$ & & & \\
\quad WFX$_2$ & 0.145 & 0.074 & 0.063 \\
\quad WFY$_2$ & 0.120 & 0.061 & 0.079 \\
WFY$_3$ $\sim$ & & & \\
\quad WFY$_2$ & 0.270 & 0.065 & 0.000 \\
\quad WFX$_2$ & 0.210 & 0.066 & 0.001 \\
WFX$_4$ $\sim$ & & & \\
\quad WFX$_3$ & 0.042 & 0.073 & 0.369 \\
\quad WFY$_3$ & 0.041 & 0.057 & 0.333 \\
WFY$_4$ $\sim$ & & & \\
\quad WFY$_3$ & 0.241 & 0.070 & 0.001 \\
\quad WFX$_3$ & 0.086 & 0.075 & 0.216 \\
WFX$_5$ $\sim$ & & & \\
\quad WFX$_4$ & 0.189 & 0.043 & 0.002 \\
\quad WFY$_4$ & 0.130 & 0.068 & 0.055 \\
WFY$_5$ $\sim$ & & & \\
\quad WFY$_4$ & 0.167 & 0.080 & 0.037 \\
\quad WFX$_4$ & 0.242 & 0.037 & 0.000 \\
WFX$_4$ $\sim$ & & & \\
\quad M & 0.561 & 0.039 & 0.000 \\
\quad WFX$_2$ & 0.455 & 0.071 & 0.000 \\
M $\sim$ & & & \\
\quad x$_{11}$ & 0.629 & 0.019 & 0.000 \\
\hline
\midrule
\multicolumn{4}{l}{Model Fit} \\
$\chi^2$ & 200.318 & & \\
$\mathit{df}$  & 187 & & \\
CFI & 0.999 & & \\
NFI & 0.981 & & \\
TLI & 0.998 & & \\
RMSEA & 0.008 & & \\
\hline
\bottomrule
\end{tabular}
\label{tab:table_a7}
\caption*{\footnotesize Note: All coefficients are standardized.}
\end{table}

\clearpage 

\section*{Cross-Lagged Panel Model (CLPM)}

We construct a CLPM model to test the 2SLW approach, following the same simulation procedure described earlier in the paper. In the population model, AR coefficients are set to 0.5, CL coefficients to 0.2, and residual correlations within time points to 0.3. Before proceeding with the analysis, it is essential to validate the baseline CLPM population model. This baseline model serves as the foundation for generating variants to test the 2SLW approach. The baseline structure, excluding the dashed lines, is illustrated in Figure A3.

\renewcommand{\arraystretch}{0.75}
\begin{table}[htbp]
\centering
\caption{Monte Carlo and Asymptotic Results for CLPM}
\vspace{-0.3cm}
\begin{tabular}{c c c c c c c c}
\hline
N & $\chi^2$ & SD & $P$-value & Rej Rate & NFI & CFI & RMSEA \\
\hline
100   & 13.060 & 5.332 & 0.444 & 0.076 & 0.981 & 1.000 & 0.000 \\
200   & 12.640 & 5.035 & 0.463 & 0.060 & 0.995 & 1.000 & 0.000 \\
300   & 11.923 & 4.850 & 0.501 & 0.046 & 0.994 & 1.000 & 0.000 \\
500   & 12.171 & 5.076 & 0.494 & 0.058 & 0.997 & 1.000 & 0.000 \\
800   & 11.746 & 4.906 & 0.521 & 0.064 & 0.995 & 0.998 & 0.027 \\
1000  & 12.000 & 5.085 & 0.501 & 0.058 & 0.997 & 1.000 & 0.012 \\
5000  & 11.794 & 4.982 & 0.513 & 0.036 & 1.000 & 1.000 & 0.000 \\
10000 & 12.130 & 4.794 & 0.490 & 0.036 & 1.000 & 1.000 & 0.006 \\
\hline
\midrule
\end{tabular}
\label{tab:table_a8}
\caption*{\footnotesize Note: Theoretically expected mean $\chi^2=12$, SD= 4.898, $p$-value=0.5, rejection rate=0.05.}
\end{table}

As shown in Table~\ref{tab:table_a8}, the $\chi^2$ test statistics are close to the degrees of freedom ($\text{df} = 12$), with a standard deviation of 4.898, an average $p$-value of 0.05, and an average rejection rate of 0.05. The fit indices, NFI and CFI, are near 1.0, while RMSEA approaches 0. These results indicate that the population model adheres to asymptotic properties and it is well specified.

\begin{figure}[H]
    \centering
    \caption{Diagram of CLPM (Correlation)}
    \includegraphics[width=0.7\textwidth]{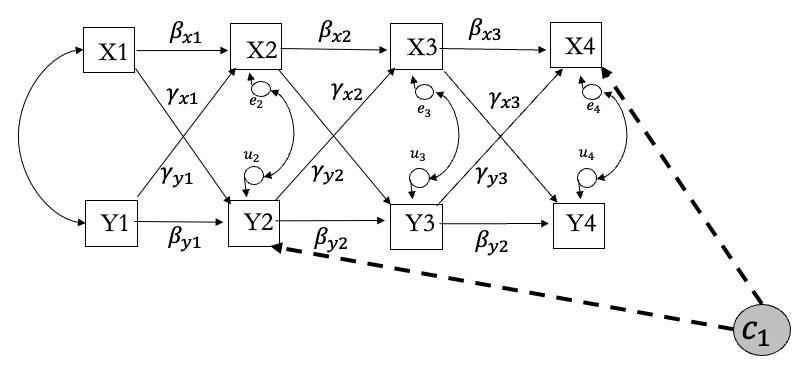} 
    \label{fig:A4}
\end{figure}

\renewcommand{\arraystretch}{0.75}
\begin{table}[H]
\centering
\caption{Test Result for CLPM (Correlation)}
\vspace{-0.3cm}
\begin{tabular}{lllll}
\hline
 & LM & EPC & Wald Test & $P$-value \\
\hline
\rowcolor{gray!20} X$_4$ $\sim\sim$ Y$_2$ & 282.919 & 0.426 & 217.073 & 0.000 \\
X$_4$ $\sim\sim$ Y$_3$ & 210.982 & -0.440 & 0.507 & 0.476 \\
X$_4$ $\sim$ Y$_2$ & 161.480 & 0.411 & 0.282 & 0.595 \\
X$_2$ $\sim\sim$ X$_4$ & 64.346 & -0.140 & 0.695 & 0.405 \\
Y$_2$ $\sim\sim$ Y$_4$ & 60.740 & -0.149 & 0.021 & 0.884 \\
Y$_4$ $\sim$ Y$_2$ & 37.786 & -0.150 & 0.240 & 0.624 \\
X$_2$ $\sim\sim$ Y$_4$ & 10.857 & 0.043 & 0.140 & 0.708 \\
\rowcolor{gray!20} Y$_4$ $\sim\sim$ Y$_1$ & 3.925 & 0.042 & 3.940 & 0.047 \\
\hline
\bottomrule
\end{tabular}
\label{tab:table_a9}
\end{table}

\noindent Table~\ref{tab:table_a9} shows that the 2SLW approach successfully identified the correlation between X$_4$ and Y$_2$. However, it also flagged a correlation between Y$_4$ and Y$_1$, which was only marginally significant. After including this parameter in the model, the coefficient remained statistically insignificant, suggesting that—despite being identified by the 2SLW—it lacks substantive meaning. Therefore, we decided to exclude it from the final model.

\renewcommand{\arraystretch}{0.75}
\begin{table}[htbp]
\centering
\caption{AR \& CL Effects for CLPM (Correlation)}
\vspace{-0.3cm}
\begin{tabular}{lccc}
\hline
 & Estimate & Std. Err & $P$-value \\
\hline
X$_2$ $\sim$ X$_1$ & 0.591 & 0.022 & 0.000 \\
X$_3$ $\sim$ X$_2$ & 0.483 & 0.030 & 0.000 \\
X$_4$ $\sim$ X$_3$ & 0.407 & 0.042 & 0.000 \\
Y$_2$ $\sim$ Y$_1$ & 0.391 & 0.027 & 0.000 \\
Y$_3$ $\sim$ Y$_2$ & 0.550 & 0.024 & 0.000 \\
Y$_4$ $\sim$ Y$_3$ & 0.504 & 0.035 & 0.000 \\
Y$_2$ $\sim$ X$_1$ & 0.191 & 0.027 & 0.000 \\
Y$_3$ $\sim$ X$_2$ & 0.184 & 0.029 & 0.000 \\
Y$_4$ $\sim$ X$_3$ & 0.193 & 0.035 & 0.000 \\
X$_2$ $\sim$ Y$_1$ & 0.160 & 0.023 & 0.000 \\
X$_3$ $\sim$ Y$_2$ & 0.250 & 0.025 & 0.000 \\
X$_4$ $\sim$ Y$_3$ & 0.124 & 0.044 & 0.000 \\
X$_4$ $\sim$ Y$_2$ & 0.471 & 0.029 & 0.000 \\
\hline
\midrule
\multicolumn{4}{l}{Model Fit} \\
\hline
$\chi^2$ & 6.826 &  &  \\
$\mathit{df}$  & 11 &  &  \\
CFI & 1.000 &  &  \\
NFI & 0.998 &  &  \\
TLI & 1.002 &  &  \\
RMSEA & 0.000 &  &  \\
\hline
\bottomrule
\end{tabular}
\label{tab:table_a10}
\end{table}

Table~\ref{tab:table_a10} shows that after including the confounder, the model became well specified. The autoregressive and cross-lagged coefficient estimates closely matched those of the population model. Moreover, the $\chi^2$ test statistic was 6.826 with 11 degrees of freedom, and the CFI, NFI, and TLI were all close to 1.00, with an RMSEA of zero. Together, these results confirm that the 2SLW approach successfully identified the unmeasured confounder.

\section*{Cross-Lagged Panel Model—Direct Effect}

\begin{figure}[H]
    \centering
    \caption{Diagram of CLPM (Direct Effect)}
    \includegraphics[width=0.7\textwidth]{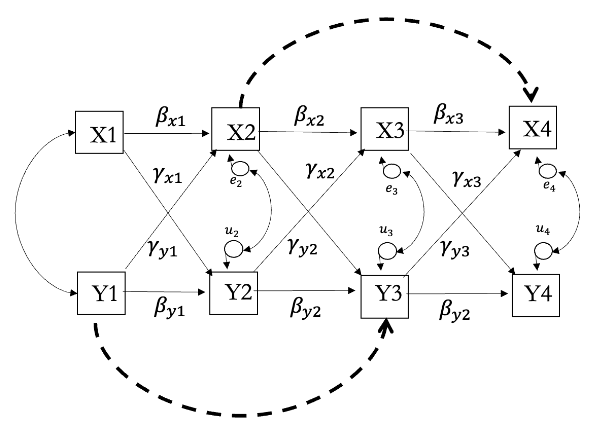} 
    \label{fig:figure_a4}
\end{figure}

In Figure~\ref{fig:figure_a4}, we create two direct effects in the CLPM, connecting $X_2$ and $X_4$, as well as $Y_1$ and $Y_3$. Table A8 presents the 2SLW results, which efficiently identified the unmeasured confounders: the LM test flagged potential parameters, and the Wald test confirmed the correct ones.

\renewcommand{\arraystretch}{0.75}
\begin{table}[htbp]
\centering
\caption{Test Result for CLPM (Direct Effect)}
\vspace{-0.3cm}
\begin{tabular}{llllll}
\hline
 & Path & LM & EPC & Wald Test & $P$-value \\
\hline
\rowcolor{gray!20} Y$_3$ & $\sim$ Y$_1$ & 477.388 & 0.639 & 914.578 & 0.000 \\
Y$_3$ & $\sim\sim$ Y$_1$ & 451.092 & 0.590 & 0.171 & 0.679 \\
\rowcolor{gray!20} X$_4$ & $\sim$ X$_2$ & 377.715 & 0.653 & 605.803 & 0.000 \\
X$_2$ & $\sim\sim$ X$_4$ & 174.765 & 0.187 & 0.600 & 0.439 \\
X$_3$ & $\sim$ Y$_1$ & 99.274 & -0.239 & 0.418 & 0.518 \\
X$_3$ & $\sim\sim$ Y$_1$ & 96.026 & -0.224 & 0.351 & 0.554 \\
Y$_4$ & $\sim$ X$_2$ & 94.751 & -0.276 & 0.333 & 0.564 \\
X$_4$ & $\sim$ X$_1$ & 77.505 & 0.223 & 0.014 & 0.905 \\
X$_4$ & $\sim\sim$ X$_1$ & 77.042 & 0.196 & 0.318 & 0.573 \\
Y$_4$ & $\sim\sim$ X$_1$ & 23.087 & -0.091 & 0.306 & 0.580 \\
Y$_4$ & $\sim$ Y$_2$ & 22.706 & -0.157 & 0.534 & 0.465 \\
\hline
\bottomrule
\end{tabular}
\end{table}

\renewcommand{\arraystretch}{0.75}
\begin{table}[H]
\centering
\caption{AR \& CL Effects for CLPM (Direct Effect)}
\vspace{-0.3cm}
\begin{tabular}{lccc}
\toprule
Path &Estimate & Std. Err & $P$-value \\
\midrule
X$_2$ $\sim$ X$_1$ & 0.536 & 0.024 & 0.000 \\
X$_3$ $\sim$ X$_2$ & 0.522 & 0.033 & 0.000 \\
X$_4$ $\sim$ X$_3$ & 0.366 & 0.035 & 0.000 \\
Y$_2$ $\sim$ Y$_1$ & 0.535 & 0.022 & 0.000 \\
Y$_3$ $\sim$ Y$_2$ & 0.354 & 0.036 & 0.000 \\
Y$_4$ $\sim$ Y$_3$ & 0.607 & 0.023 & 0.000 \\
Y$_2$ $\sim$ X$_1$ & 0.190 & 0.023 & 0.000 \\
Y$_3$ $\sim$ X$_2$ & 0.151 & 0.031 & 0.000 \\
Y$_4$ $\sim$ X$_3$ & 0.159 & 0.032 & 0.000 \\
X$_2$ $\sim$ Y$_1$ & 0.204 & 0.023 & 0.000 \\
X$_3$ $\sim$ Y$_2$ & 0.185 & 0.034 & 0.000 \\
X$_4$ $\sim$ Y$_3$ & 0.153 & 0.024 & 0.000 \\
Y$_3$ $\sim$ Y$_1$ & 0.500 & 0.021 & 0.000 \\
X$_4$ $\sim$ X$_2$ & 0.443 & 0.027 & 0.000 \\
\hline
\midrule

\multicolumn{4}{l}{Model Fit} \\
\midrule
$\chi^2$ & 3.546 &  &  \\
$\mathit{df}$        & 10    &  &  \\
CFI      & 1.000 &  &  \\
NFI      & 0.999 &  &  \\
TLI      & 1.003 &  &  \\
RMSEA    & 0.000 &  &  \\
\hline
\bottomrule
\end{tabular}
\label{tab:table_a11}
\caption*{\footnotesize Note: All coefficients are standardized.}
\end{table}

Table~\ref{tab:table_a11} shows that, after incorporating the suggested unmeasured confounders into the original model, the $\chi^2$ statistic and fit indices indicate excellent fit. Moreover, the AR and CL coefficients closely match those of the population model.

\section*{Cross-Lagged Panel Model---Mediation}

\begin{figure}[H]
    \centering
    \caption{Diagram of CLPM (Mediation)}
    \includegraphics[width=0.7\textwidth]{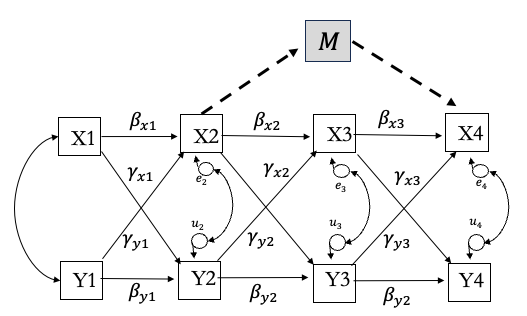} 
    \label{fig:A6}
\end{figure}

To test whether the 2SLW approach can identify a mediation path in the CLPM, we specify the model shown in Figure~\ref{fig:A6}, where $M$ serves as the mediator between $X_2$ and $X_4$. The model includes both the direct effect of $X_2$ on $X_4$, controlling for $M$, and the indirect effect of $X_2$ on $X_4$ through $M$. To construct this mediation structure, we regress $M$ on $x_1$, inducing a high correlation between the two. 

Table~\ref{tab:table_a12} shows that the 2SLW approach identifies four unmeasured confounders. Specifically, $X_4$ is predicted by both $X_2$ and $M$, as well as by $x_1$. Because $M$ is regressed on $x_1$, their LM test statistics are identical, and the Wald test reveals a multicollinearity issue between them. Therefore, we disregard the path $Y_4 \sim x_1$ in subsequent analyses. Additionally, the 2SLW approach identifies $Y_4 \sim x_1$ as statistically significant. 

To validate this suggested confounder, we include it in the revised model. As shown in Table~\ref{tab:table_a12}, the coefficient for $Y_4 \sim x_1$ is small and not statistically significant, indicating it carries little explanatory value. We recommend excluding this parameter from further analysis.

\renewcommand{\arraystretch}{0.75}
\begin{table}[H]
\centering
\caption{AR \& CL Effects for CLPM (Mediation)}
\vspace{-0.3cm}
\begin{tabular}{lccc}
\hline
Path & Estimate & Std. Err & $P$-value \\
\hline
X$_2$ $\sim$ X$_1$ & 0.548 & 0.023 & 0.000 \\
X$_3$ $\sim$ X$_2$ & 0.507 & 0.033 & 0.000 \\
X$_4$ $\sim$ X$_3$ & 0.284 & 0.038 & 0.000 \\
Y$_2$ $\sim$ Y$_1$ & 0.521 & 0.023 & 0.000 \\
Y$_3$ $\sim$ Y$_2$ & 0.489 & 0.035 & 0.000 \\
Y$_4$ $\sim$ Y$_3$ & 0.528 & 0.035 & 0.000 \\
Y$_2$ $\sim$ X$_1$ & 0.237 & 0.023 & 0.000 \\
Y$_3$ $\sim$ X$_2$ & 0.204 & 0.034 & 0.000 \\
Y$_4$ $\sim$ X$_3$ & 0.125 & 0.035 & 0.001 \\
X$_2$ $\sim$ Y$_1$ & 0.194 & 0.023 & 0.000 \\
X$_3$ $\sim$ Y$_2$ & 0.191 & 0.034 & 0.000 \\
X$_4$ $\sim$ Y$_3$ & 0.155 & 0.035 & 0.000 \\
c$_1$ $\sim$ x$_1$ & 0.923 & 0.007 & 0.000 \\
X$_4$ $\sim$ X$_2$ & 0.450 & 0.026 & 0.000 \\
M & 0.314 & 0.038 & 0.000 \\
Y$_4$ $\sim$ x$_1$ & 0.042 & 0.020 & 0.074 \\
\hline
\midrule
\multicolumn{4}{l}{Model Fit} \\
\midrule

$\chi^2$ & 28.972 & & \\
$\mathit{df}$        & 26     & & \\
CFI      & 1.000  & & \\
NFI      & 0.996  & & \\
TLI      & 0.999  & & \\
RMSEA    & 0.011  & & \\
\hline
\bottomrule
\end{tabular}
\label{tab:table_a12}
\caption*{\footnotesize Note: All coefficients are standardized.}
\end{table}

Table~\ref{tab:table_a12} shows that the model is well specified, as indicated by a $\chi^2$ statistic of 28.972 with $\textit{df}=26$. All fit indices (CFI, NFI, TLI, and RMSEA) fall within acceptable ranges. Moreover, all AR and CL coefficients are statistically significant, except for the path from X1 to Y4. Additionally, all parameter estimates are close to those of the population model, indicating that the 2SLW approach successfully identified the unmeasured mediator.

\end{document}